\documentclass[twocolumn]{aastex62}

\submitjournal{ApJ}
\usepackage[all]{hypcap} 
\usepackage{tabularx, graphicx}

\usepackage{bm}
\usepackage{amsmath}
\usepackage{color}

\newcommand{\lenstool}{{\tt{Lenstool}}}
\newcommand{\LENSTOOL}{{\tt{Lenstool}}}

\newcommand{\zs}{z_\mathrm{S}}
\newcommand{\zl}{z_\mathrm{L}}

\newcommand{\eER}{\rm{e}\theta_{\rm{E}}}

\newcommand{\MSHM}{M_{\rm SHM}}
\newcommand{\MSIM}{M_{\rm SIM}}

\newcommand{\rmsi}{{\it rmsi}}
\newcommand{\AllScatter}{{$8.52$\%}}
\newcommand{\AllBias}{{$0.90$\%}}
\newcommand{\GSHMScatter}{{$3.26$\%}}
\newcommand{\GSHMBias}{{$0.34$\%}}
\newcommand{\NOZSSHMScatter}{{$9.85$\%}}
\newcommand{\NOZSSHMBias}{{$-7.22$\%}}
\newcommand{\SLCONFSHMScatter}{{$3.88$\%}}
\newcommand{\SLCONFSHMBias}{{$0.84$\%}}

\begin{document}
\title{Core Mass Estimates in Strong Lensing Galaxy Clusters Using a Single-Halo Lens Model}

\author[0000-0002-7868-9827]{J. D. Remolina Gonz\'{a}lez}
\affiliation{Department of Astronomy, University of Michigan, 1085 S. University Ave, Ann Arbor, MI 48109, USA}
\email{jremolin@umich.edu}

\author[0000-0002-7559-0864]{K. Sharon}
\affiliation{Department of Astronomy, University of Michigan, 1085 S. University Ave, Ann Arbor, MI 48109, USA}

\author[0000-0001-6800-7389]{N. Li}
\affil{CAS, Key Laboratory of Space Astronomy and Technology, National Astronomical Observatories, A20 Datun Road, Chaoyang District, Beijing 100012, People’s Republic of China}
\affil{School of Physics and Astronomy, Nottingham University, University Park, Nottingham NG7 2RD, UK}

\author[0000-0003-3266-2001]{G. Mahler}
\affil{Department of Astronomy, University of Michigan, 1085 S. University Ave, Ann Arbor, MI 48109, USA}
\affil{Centre for Extragalactic Astronomy, Department of Physics, Durham University, Durham DH1 3LE, UK}
\affil{Institute for Computational Cosmology, Durham University, South Road, Durham DH1 3LE, UK}

\author[0000-0001-7665-5079]{L. E. Bleem}
\affil{Argonne National Laboratory, High-Energy Physics Division, Argonne, IL 60439}
\affil{Kavli Institute for Cosmological Physics, University of Chicago, 5640 South Ellis Avenue, Chicago, IL 60637, USA}

\author[0000-0003-1370-5010]{M. Gladders}
\affil{Department of Astronomy and Astrophysics, University of Chicago, 5640 South Ellis Avenue, Chicago, IL 60637, USA}
\affil{Kavli Institute for Cosmological Physics, University of Chicago, 5640 South Ellis Avenue, Chicago, IL 60637, USA}

\author[0000-0003-3791-2647]{A. Niemiec}
\affil{Department of Astronomy, University of Michigan, 1085 S. University Ave, Ann Arbor, MI 48109, USA}
\affil{Centre for Extragalactic Astronomy, Department of Physics, Durham University, Durham DH1 3LE, UK}
\affil{Institute for Computational Cosmology, Durham University, South Road, Durham DH1 3LE, UK}

\begin{abstract}

The core mass of galaxy clusters is an important probe of structure formation. Here, we evaluate the use of a Single-Halo model (SHM) as an efficient method to estimate the strong lensing cluster core mass, testing it with ray-traced images from the ‘Outer Rim’ simulation. 
Unlike detailed lens models, the SHM represents the cluster mass distribution with a single halo and can be automatically generated from the measured lensing constraints. 
We find that the projected core mass estimated with this method, $\MSHM$,  has a scatter of \AllScatter\ and a bias of \AllBias\ compared to the ``true'' mass within the same aperture.  
Our analysis shows no systematic correlation between the scatter or bias and the lens-source system properties. The bias and scatter can be reduced to \GSHMScatter\ and \GSHMBias, respectively, by excluding models that fail a visual inspection test.  
We find that the SHM success depends on the lensing geometry, with single giant arc configurations accounting for most of the failed cases due to their limiting constraining power. When excluding such cases, we measure a scatter and bias of \SLCONFSHMScatter\ and \SLCONFSHMBias, respectively.
Finally, we find that when the source redshift is unknown, the model-predicted redshifts are overestimated, and the $\MSHM$ is underestimated by a few percent, highlighting the importance of securing spectroscopic redshifts of background sources. 
Our analysis provides a quantitative characterization of $\MSHM$, enabling its efficient use as a tool to estimate the strong lensing cluster core masses in the large samples, expected from current and future surveys.

\end{abstract}



\section{Introduction} 
\label{sec:intro}

Harbored at the high-density knots of the cosmic web, galaxy clusters trace the large-scale structure formation of the universe, making them valuable cosmological laboratories (see reviews by \citealt{Allen:11} and \citealt{Mantz:14}). 
Their mass function, which connects their observational properties to the underlying cosmology (e.g., \citealt{Jenkins:01, Evrard:02, Corless:09, Pratt:19,Bocquet:20}), is one of the ensemble properties that cluster-based cosmological studies are pursuing. 
However, the efficacy of cluster-based cosmological studies is sensitive to sample size and selection function (e.g., \citealt{Hu:03b, Khedekar:13, Bocquet:19}) and requires a good understanding of the inherent systematic errors in the mass estimate due to the observed astrophysical properties \citep{Evrard:02, Allen:11, Huterer:18}. 
Other cluster properties predicted by cosmological simulations include the radial profiles and concentrations of dark matter halos (\citealt{Duffy:08, Meneghetti:14, Child:18}), and can be directly tested with observations (e.g., \citealt{Oguri:12,Merten:15}).
An accurate measurement of the mass profile slope of galaxy clusters requires mass proxies that are sensitive to the total cluster mass, as well as mass proxies whose resolution is high enough to probe the innermost hundreds of parsecs. 
 
Gravitational lensing probes the total (dark and baryonic) matter distribution, independent of baryonic physics and cluster dynamical state. Strong gravitational lensing (SL) has the highest resolution at the core of galaxy clusters, where the strong lensing evidence is present; Weak lensing (WL) gives an accurate measurement of the total mass at large cluster-centric radii. By combining the mass estimate from SL at the core with a mass estimate at large scales from WL or other mass proxies, we can constrain the mass distribution profile from the core to the outskirts, and measure profile parameters such as the concentration of the galaxy cluster (e.g., \citealt{Gralla:11, Oguri:12, Merten:15}). Tension between the observations and theoretical expectation of the mass distribution profile of SL galaxy clusters has been reported (e.g., \citealt{Broadhurst:08, Gonzalez:12, Meneghetti:13}), however, these studies are limited by small samples and complicated selection functions.

Thousands of SL galaxy clusters are being discovered with current and up-coming large surveys, covering a broad wavelength range, detecting clusters out to $z \sim 2$, and addressing challenges due to small sample sizes. These include cluster surveys based on observations with the South Pole Telescope (SPT; SPT-3G, \citealt{Benson:14}; SPT-SZ 2500 deg$^2$, \citealt{Bleem:15, Bocquet:19}; SPT-Pol 100 deg$^2$, \citealt{HuangN:20}; SPT-ECS, \citealt{Bleem:20}), the Atacama Cosmological Telescope (ACT; \citealt{Marriage:11,Hilton:18}), the Cerro Chajnantor Atacama Telescope (CCAT; \citealt{Mittal:18}), Euclid \citep{Laureijs:11,Amendola:18}, Vera Rubin Observatory Legacy Survey of Space and Time (LSST, \citealt{LSST:17}), and eROSITA \citep{Pillepich:18}. We expect that hundreds of the newly discovered clusters will be strong lenses \citep{LSST:09}. With an order of magnitude increase in sample sizes, an efficient and accurate method will be required in order to measure the mass at the cores of the SL clusters in a timely manner.

Strong lensing-based measurements of the mass distribution at the cores of galaxy clusters typically rely on detailed lensing analyses. A detailed lens model of a cluster with rich strong lensing evidence (e.g., the Frontier Fields; \citealt{Lotz:17}) can have a high level of complexity requiring a large number of constraints, extensive follow-up observations, computational resources, and multiple iterations to be finalized (e.g., \citealt{Johnson:14, Zitrin:14, Diego:16, Kawamata:16, Lotz:17, Strait:18, Lagattuta:19, Sebesta:19, Raney:20a}). Due to the limited resources and small number of lensing constraints, which is typical for all but the most massive strong lensing clusters (e.g., \citealt{Sharon:20}), 
there is a need to investigate efficient methods to estimate the mass at the core of SL galaxy clusters. \citet{Remolina:20} presented an evaluation of the mass enclosed by the Einstein radius as a zeroth-order method to estimate the mass at the core of galaxy clusters. The limiting factor when using the Einstein radius to estimate the core mass is the assumption of spherical symmetry inherent to this method. In this paper, we investigate a higher complexity first-order method, which is more complex than the mass enclosed by the Einstein radius, but not as expensive as computing a detailed lens model.

The goal of this paper is to evaluate the use of the Single-Halo model (SHM) as an efficient method to measure the mass at the core of SL galaxy clusters. We measure the scatter and bias in the mass estimate, establish limitations in the use of the SHM, and explore dependence of the scatter on the properties of the model and the lens-source system. We use the state-of-the-art `Outer Rim' simulation run \citep{Heitmann:19}, which facilitates a robust statistical analysis that is representative of the universe.

This paper is organized as follows. In \S\ref{sec:lens_modeling}, we describe the lensing algorithm used in our analysis, \LENSTOOL, summarize the procedure employed in detailed lens models, and present the Single-Halo model. In \S\ref{sec:data}, we describe the `Outer Rim' simulation and detail the simulated sample used in our analysis. In \S\ref{sec:methodology}, we describe the identification of constraints for the SHM, compute the SHM, and measure the aperture mass enclosed within the effective Einstein radius, $\MSHM$. In \S\ref{sec:results}, we measure the bias and scatter of $\MSHM$, in comparison to the true mass. In \S\ref{sec:no_zs_and_lens_config}, we investigate the effects on the SHM of an unknown background source redshift, the lensing geometry of the arc, and addition of a second multiply-imaged source. Last, we present our conclusion and summary of the evaluation of the Single-Halo models as a mass estimate at the core of galaxy clusters in \S\ref{sec:conclusion}.

In our analysis, we adopt a \textit{WMAP}-7 \citep{Komatsu:11} flat $\Lambda$CDM cosmology as in the `Outer Rim' simulation: $\Omega_{\Lambda} = 0.735$, $\Omega_{M} = 0.265$, and $h = 0.71$. Masses reported in terms of M$_{\mathrm{\Delta c}}$, are defined as the mass enclosed within a radius at which the average density is $\Delta$ times the critical density of the universe at the cluster redshift.


\section{Background: Lens Modeling} 
\label{sec:lens_modeling}

Strong lens modeling analyses use the positional and redshift measurements of lensed galaxies (arcs) as constraints to model the underlying mass distribution. We use the publicly available lens modeling algorithm \LENSTOOL\ \citep{Jullo:07}, which has been widely used (e.g., \citealt{Johnson:14, Cerny:18, Paterno-Mahler:18, Lagattuta:19, Jauzac:20, Mahler:20, Sharon:20}) and its results are similar to other parametric models \citep{Meneghetti:17, Priewe:17, Remolina:18, Raney:20b}. \LENSTOOL\ uses a Monte Carlo Markov Chain (MCMC) method to explore the parameter space, identify the best fit values, and estimate the statistical uncertainties in the model. To characterize the mass density distribution we use a parameterized dual pseudo-isothermal ellipsoid (dPIE, \citealt{Eliasdottir:07}) with seven parameters: position ($\alpha$ and $\delta$), ellipticity ($\epsilon = (a^2 - b^2)/(a^2 + b^2)$), where $a$ and $b$ are the semi-major and semi-minor axis respectively), position angle ($\theta$), velocity dispersion ($\sigma$), core radius (R$_{core}$), and truncating radius (R$_{cut}$). We fix the truncating radius (R$_{cut}$) to $1500$ kpc, since it is far beyond the lensing region, and cannot be constrained using the strong lensing evidence. We note that this range resembles the splashback radius (e.g., \citealt{Umetsu:17,Shin:19}).
In the next subsections we describe the difference between ``detailed'' and Single-Halo models, and describe the selection of constraints and priors used for the lens modeling procedure.

\subsection{Detailed Lens Models} 
\label{subsec:dlm}

For in-depth description on the commonly used procedures in detailed parametric lens modeling, we refer the reader to \citet{Verdugo:11} and \citet{Richard:11}, and examples by \citet{Mahler:18, Mahler:20, Lagattuta:19}; and \citet{Sharon:20}. Detailed lens models use the galaxy cluster redshift, and the position and redshift of the arcs as constraints. One or more large cluster-scale profile(s) represent the dark matter halo(s) of the cluster and correlated structure as needed, and multiple galaxy-scale halos represent the galaxy cluster members mass contribution. The galaxy-scale potentials positional parameters are usually fixed to their observed values and a parameterized mass-luminosity relation is used to set or fit the other parameters. The brightest cluster galaxy (BCG) may be modeled with a separate halo as we do not expect BCGs to follow the same mass-luminosity relation as the rest of the cluster members. 

Compared to the Single-Halo models that will be introduced in the next section, detailed lens models can be highly complex. The complexity adds the flexibility needed in order to trace the substructure in the form of multiple dark matter halos, filaments, contributions from cluster-member galaxies, and in some cases uncorrelated structure along the line-of-sight. The versatility of these models has been shown to be a successful tool for studying a broad range of sciences including cosmology, galaxy clusters physics, and the highly magnified background universe (e.g., \citealt{Johnson:17b, Acebron:17,Gonzalez:20}. The flexibility of detailed lens models also means the models are not unique and require care in the construction and evaluation, often multiple statistical assessments to select between models (e,g., \citealt{Acebron:17, Paterno-Mahler:18, Lagattuta:19, Mahler:20}).  

Detailed lens models for galaxy clusters with rich strong lensing evidence require extensive follow-up observations, computational resources, and multiple iterations of the modeling process. The high complexity of the models relies on a large number of free parameters, requiring a large number of constraints, i.e., multiply-imaged lensed galaxies, whose availability becomes a limiting factor in the modeling process.

\subsection{Single-Halo Models} 
\label{subsec:shm}

Single-Halo models are similar to their detailed counterparts and use the same type of constraints. The difference in the SHM modeling procedure is that the lens plane is described by a single cluster-scale dark matter halo, while all secondary halos and contribution from cluster member galaxies are neglected. The small number of parameters requires only a handful of constraints, and the model can be computed quickly and with limited human intervention.

We use the same dPIE halos described above, with six free parameters. 
We use broad priors in the parameters of the dPIE potential: $-8\farcs0 < \alpha, \delta < 8\farcs0$ ; $0.0 < \epsilon < 0.9$ ;  $0^{\circ}  < \theta <180^{\circ}$ ; $50$ kpc $<$ R$_{core}$ $< 150$ kpc; and $500$ km/s $< \sigma < 1500$ km/s.

The outputs of the lens models include the projected mass distribution ($\Sigma$), convergence ($\kappa$), shear ($\gamma$), magnification ($\mu$), critical curves, and predicted location of multiple-images. The tangential critical curve (TCC) and radial critical curve (RCC) are the theoretical lines of infinite magnification and name the primary direction along which images (arcs) are magnified. The magnification in the tangential direction is computed as follows: $\mu_{t}^{-1} = 1 - \kappa - \gamma$. In this analysis, we measure the aperture mass enclosed by the effective Einstein radius ($\eER$), defined as the radius of a circle with the same area as the area enclosed by the tangential critical curves.


\section{Simulated Data:} 
\label{sec:data}

To evaluate the SHM method, we use the state-of-the-art, large volume, high-mass-resolution, N-body simulation `Outer Rim' \citep{Heitmann:19} with the HACC framework \citep{Habib:16}. The simulation was carried out at the  Blue  Gene/Q system at Argonne National Laboratory. 
The large size simulation box (L = $3000$ Mpc h$^{-1}$ on the side) allows for many massive halos in the redshift range of interest ($z \sim 0.1 - 0.7$) with detailed projected mass distribution profiles representative of the universe.

The `Outer Rim' simulation has been used to study the dark matter halo profiles of galaxy groups and clusters \citep{Child:18}, evaluate the effects on lensing due to line-of-sight (LOS) structure \citep{Li:19}, and to construct realistic strong lensing ray-traced simulated images \citep{Li:16}. The simulation does not include the baryonic component; while baryons represent a small portion of the mass content of the galaxy cluster, studies have shown that the baryonic component has   
non-zero effects on the mass distribution and the lensing potential. For example, the concentration of dark matter halos is higher when baryons are included in the simulation  \citep{Meneghetti:03, Wambsganss:04, Oguri:06, Hilbert:07, Hilbert:08, Wambsganss:08, Oguri:09}.  The light due to the baryonic component is also not depicted in the simulated images, i.e., the diffused light from intracluster medium, and stellar population of cluster member galaxies. Fully accounting for these baryonic effects awaits for simulations that include baryonic physics in large cosmological boxes.

The galaxy cluster halos used in the analysis were identified using a friends-of-friends algorithm with linking length of b $= 0.168$ and the surface density was computed using a density estimator. \citet{Rangel:16} showed that the high mass resolution is robust enough to simulate strong lensing in halos with masses M$_{500c} > 2 \times 10^{14}$ M$_{\odot}$ h$^{-1}$. Following an SPT-like selection function, all halos with M$_{500c} > 2.1 \times 10^{14}$ M$_{\odot}$ h$^{-1}$ were selected. From this sample, the strong lenses are identified as those having an Einstein radius of at least a few arcseconds, as measured from the average convergence $\langle \kappa(\theta) \rangle = 1$.

The sample details are presented in \citet{Remolina:20} and summarized here. The sample of simulated SPT-like strong lenses is made of $74$ galaxy cluster halos spanning a redshift range of $z \sim 0.16 - 0.67$. The redshift range is similar to other strong lensing samples like that of the Sloan Giant Arc Survey (SGAS, Gladders et al, in preparation; \citealt{Sharon:20}). Future studies will extend the redshift range, $z < 1.75$, to better match surveys like the SPT-SZ 2500-Square-Degree survey \citep{Bleem:15}. We adopt the halo masses  ($\mathrm{M}_{200c}$) and concentrations ($\mathrm{c}_{200c}$) that were derived by  \citet{Child:18}.  

The lensed simulated images were created through ray-tracing using the projected mass distribution of the galaxy cluster following the procedure detailed in \citet{Li:16}. We draw redshifts for the background sources following the observed distribution of \citet{Bayliss:11} leading to a simulated source range of $z\sim 1.2$ to $z \sim 2.7$. The image plane of each cluster field is generated $5$ to $24$ times, each realization using a single redshift and unique background source location. A total of $1024$ simulated ray-traced realizations were created from the $74$ simulated SL galaxy cluster halos. The simulated images have a resolution of $0\farcs09$ per pixel and a field of view of 2048x2048 pixels. No additional noise or errors were added, as we use these simulations to investigate the most ideal case rather than creating mock observations that simulate a particular instrument. The background sources were preferentially placed in areas of high magnification, as highly magnified (total magnification $> 5$) arcs are easily detected from ground based observations. This strategy was chosen in order to mock the selection function of lensing-selected samples, in which lensing clusters were identified based on the appearance of a giant arc in visual inspection of shallow observational data (e.g., \citealt{Bayliss:11, Nord:16, Nord:20, Sharon:20, Khullar:20}). The total magnification is defined as the ratio of the area between the image-plane and source-plane of the lensed image. Only isolated halo ray-tracing is used, no structure along the line-of-sight was included. Structure along the line-of-sight is known to affect the lensing potential (e.g. \citealt{Bayliss:14, DAloisio:14, Chiviri:18, Li:19}). A statistical analysis of the line-of-sight effect and lensing systems without dominant giant arcs is left for future work. 

We use the redshift and observed image plane positions of the arcs as our constraints for the lens models. We compare the model mass we derive from the SHM method to the projected mass density from the simulation.


\section{Methodology:} 
\label{sec:methodology}

Our analysis of the simulated ray-traced lensed images is guided by the procedures used with observational data. The images are inspected one at a time to identify the multiply-imaged morphological features (emission knots) to be used as positional constraints in the lens modeling process. In the case of observational data, visual inspection is also required for spectroscopic follow-up observation of the arcs and cluster members. Here, we assume that the redshifts of the arcs and the clusters have been measured (see \S\ref{sec:data} and \citealt{Remolina:20}). In this section, we provide a description of the identification of the lensing evidence, compute SHMs, and estimate the mass at the core of the galaxy cluster within the $\eER$.

\subsection{Arc Catalog Identification}
\label{subsec:arc_flag_id}

Identical to the procedure described in \citet{Remolina:20}, we identify the lensing evidence and measure the positional constraints in the simulated lensed images. Lens modelers take advantage of the expected lensing geometry, morphology, and color information to associate sets of multiple-images of the same background source. In our analysis no color information was implemented, so we rely on the morphology and expected lensing geometry for this identification. For each line of sight, we compile a catalog including the positional locations of the arcs (including identified emission knots within each arc) and their redshift. Each identified set of $n$ multiply imaged features contributes $2n-2$ constraints.

\subsection{The SHM Procedure}
\label{subsec:computing_SHM}

One of the advantages of the SHM is that it can be automatically computed with minimal human intervention. It requires as input the cluster redshift, initial center position (e.g., the BCG), and positions and redshifts of the arcs. The best fit lens model is the one that minimizes the scatter between the observed and model-predicted positions of the arcs in the image-plane. Since the single dPIE halo has six free parameters, the SHM requires a minimum of six lensing constraints. We find that of the $1024$ simulated lensed images, $938$ have enough constraints for a SHM to be computed. We note that while this requirement is satisfied, it does not guarantee that the lens model will be fully constrained and in some cases may result in unphysical SHMs, as will be discussed in the next section.

\subsection{Assessment of the SHM Success}
\label{subsec:shm_visual_inspec}
A quick visual inspection of the resultant critical curve and model predictions with respect to the lensing evidence can provide a critical quality assessment of the Single-Halo models. We inspect each of the generated models, and find that in some cases the Single-Halo model does not reproduce the lensing configuration, and/or predicts multiple images in areas where no arcs are found. We flag these cases as ``Failed Single-Halo Models'' (F-SHM); such models would not be trusted in a typical observational analysis and would usually require a more involved lens modeling analysis, or more constraints to improve the fidelity of the models. 

\begin{figure*}
\center
\includegraphics[width=1\textwidth]{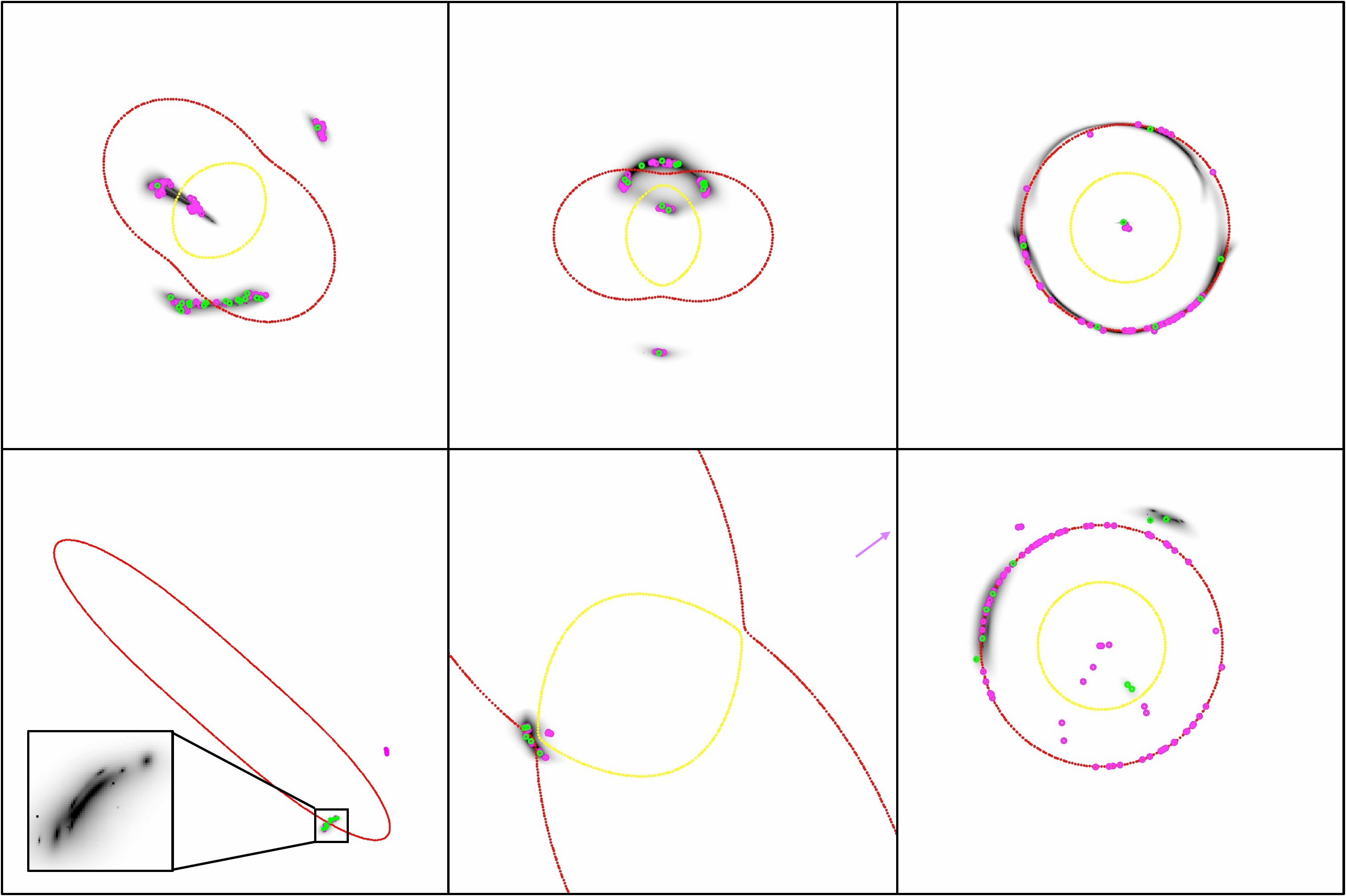}

\caption{\textsc{\textbf{Examples of SHM outputs}}, overplotted on six ray-traced lensing images.  The tangential and radial critical curves for the redshift of the background source are plotted in red and yellow, respectively. The green circles mark the constraints, and the magenta circles show the model-predicted image locations. Each image is $1.0$ arcminute (except the \textit{Bottom-Left}, which is $2.0$ arcminutes) on the side and a resolution of $0\farcs09$ per pixel. 
\textit{Top row}: representative models that pass the visual inspection test (P-SHM). Each of these models predicts lensed images at their observed locations. \textit{Bottom row:} models that fail to reproduce the lensing geometry (F-SHMs). The primary reason for rejecting these models is as follows:
\textit{Bottom-Left}: 
The lensing configuration, arc curvature, and the unrealistically high ellipticity of the SHM halo suggest that there is a contribution from a secondary mass halo, which cannot be well represented by a single halo model.
\textit{Bottom-Middle}: The SHM critical curves are extremely large leading to an unphysical mass distribution for the lensing configuration, also producing projections where no arcs are found outside the shown field of view indicated by the magenta arrow. \textit{Bottom-Right}: The SHM predicts lensed images where no arcs are found.}
\label{fig:ng_shm_ex}
\end{figure*}

We flagged $201$ out of our $938$ models as F-SHMs, leaving $737$ that pass the visual inspection (P-SHM); \autoref{fig:ng_shm_ex} shows representative examples of P-SHMs (top row) and F-SHMs (bottom row). Each of the 74 galaxy clusters still has at least one SHM that passed the visual inspection, with most having $9$ or more P-SHMs.
We note that due to the construction process of our simulated images, i.e., the background sources were preferentially placed in order to produce highly magnified images, the SHM success rate we quote here does not represent the expected success rate in the Universe; it is more tuned to resemble the success rate of modeling systems with giant arcs (e.g., \citealt{Johnson:17a,Rigby:18,Sharon:20,Remolina:21}).
We use the fail/pass distinction in \S\ref{sec:results}.

\begin{figure}
\includegraphics[width=0.5\textwidth]{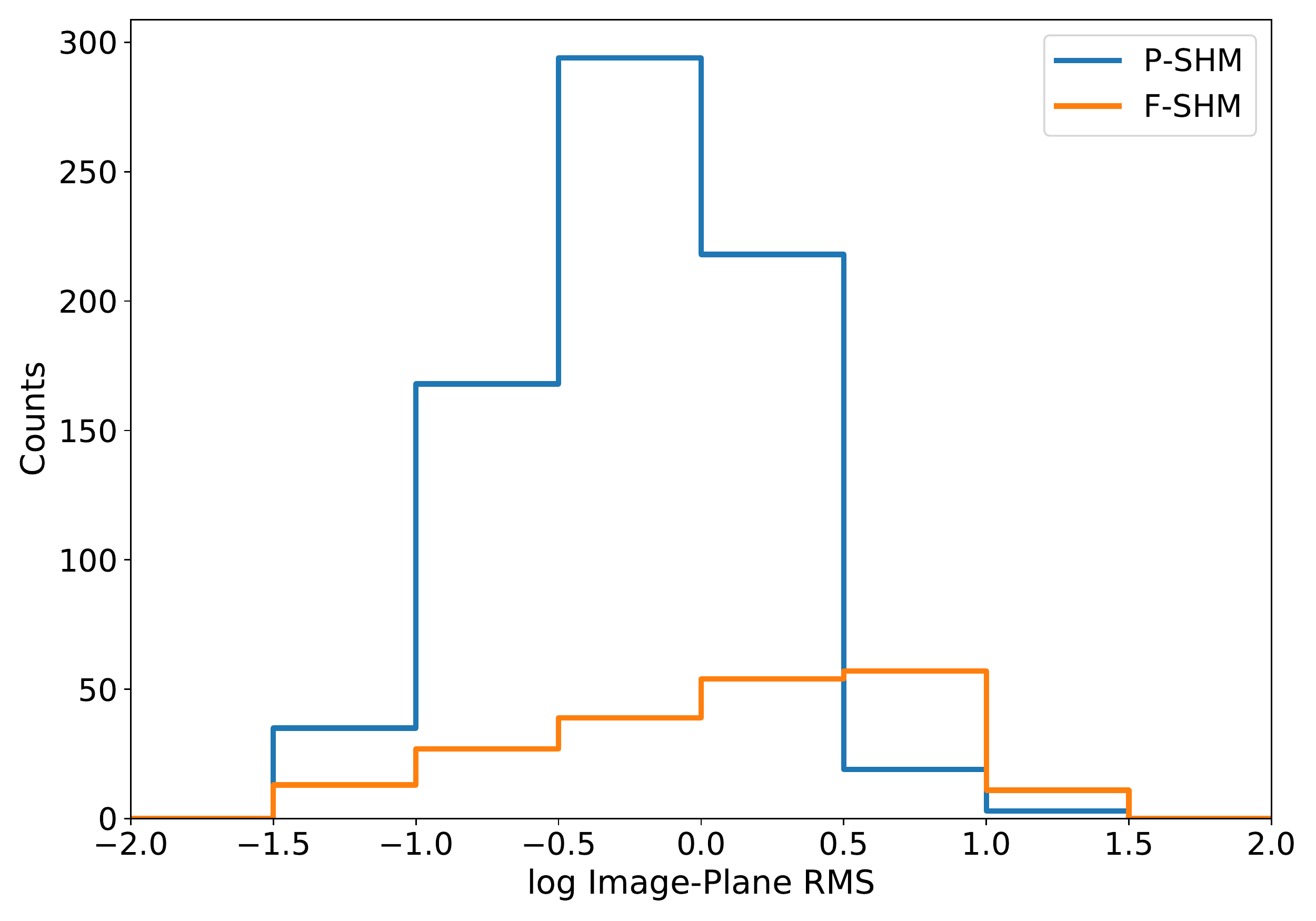}
\caption{\textsc{\textbf{Distribution of Image-Plane Root-Mean-Square (\rmsi) of the Lens Models.}} {
The distributions of models that passed (P-SHM) or failed (F-SHM) the visual inspection are shown in blue and orange, respectively. The F-SHM distribution is skewed towards higher \rmsi\ values, but both F-SHM and P-SHM can have low \rmsi\ values, making this an insufficient predictor of model quality.
}}
\label{fig:rmsi_Vs_visual_inspection}
\end{figure}

We investigate whether the image-plane root-mean-square (\rmsi) can be used as a quantitative quality indicator in lieu of a visual inspection.
The \rmsi\ is often used to determine the goodness of fit of lens models; it measures the scatter between the observed and model-predicted image-plane locations of lensed features, and in most strong lens modeling algorithms it is used in the minimization process. We find that the value of the model \rmsi\ is only a weak predictor of the quality of the SHMs. As can be seen in \autoref{fig:rmsi_Vs_visual_inspection}, while the highest bins of \rmsi\ are dominated by F-SHMs, both P-SHMs and F-SHMs span the full range of low \rmsi\ bins. This means that a low \rmsi\ is not a sufficient indicator of model quality. This finding is consistent with previous studies. In an observational analysis of 37 lensing clusters, \citet{Sharon:20} found that while the \rmsi\ serves as a good statistical indicator when comparing different lens models of the exact same system, it is not a good absolute predictor of a lens model quality. \citet{Johnson:16} used simulations to study the relationship between the \rmsi, the number of constraints, and the accuracy in recovering the mass and magnification. They found that as expected, the accuracy of the magnification and mass recovered by the lens models improved with larger numbers of constraints, however, the \rmsi\ increases with number of constraints. We therefore do not recommend relying on \rmsi\ alone in order to determine which models pass or fail.

\subsection{Aperture Mass Enclosed by the $\eER$}
\label{subsec:get_masses}

We use the projected mass distribution ($\Sigma$) from the best fit lens model to compute other outputs including the magnification ($\mu$), convergence ($\kappa$), and shear ($\gamma$). We compute the magnification in the tangential direction ($\mu_{t}$) and determine the location of the TCC ($\mu_{t} \to \infty$). Next, we measure the $\eER$ as the radius of the circle with the same area as the area enclosed by the TCC. Last, we measure the aperture mass centered at the center of the modeled dPIE dark matter halo and enclosed by the effective Einstein radius, which we denote $\MSHM$.

\subsection{Statistics}
\label{subsec:stats}

To establish a robust statistical analysis using our simulated SL sample, we weight each of the $74$ SL galaxy cluster equally. We also weight each ray-tracing realization by a factor of one over the total number of realizations with SHMs for each galaxy cluster. Then for every simulated cluster, we randomly select one $\MSHM$ representative of a ray-tracing image. We repeat this process $1000$ times for each of the 74 simulated clusters for a total of $74,000$ mass measurements used in our statistical analysis.


\section{Analysis of Results:} 
\label{sec:results}

In this section, we compare the aperture mass enclosed by the effective Einstein radius of the Single-Halo model ($\MSHM$) and the ``true'' mass enclosed within the same aperture from the simulation ($\MSIM$). We compute the scatter and bias of $\MSHM$ versus $\MSIM$ and explore whether the $\MSHM$ depends on the lens model parameters and the simulated galaxy cluster properties.

For the statistics used in this analysis see \S\ref{subsec:stats}. The scatter is computed as half the difference between the $84$th and $16$th percentiles. The bias is computed as  
\begin{equation}
    \mathrm{bias} = \mathrm{median}(\MSHM / \MSIM) - 1.
\end{equation}
 
\begin{figure}
\includegraphics[width=0.5\textwidth]{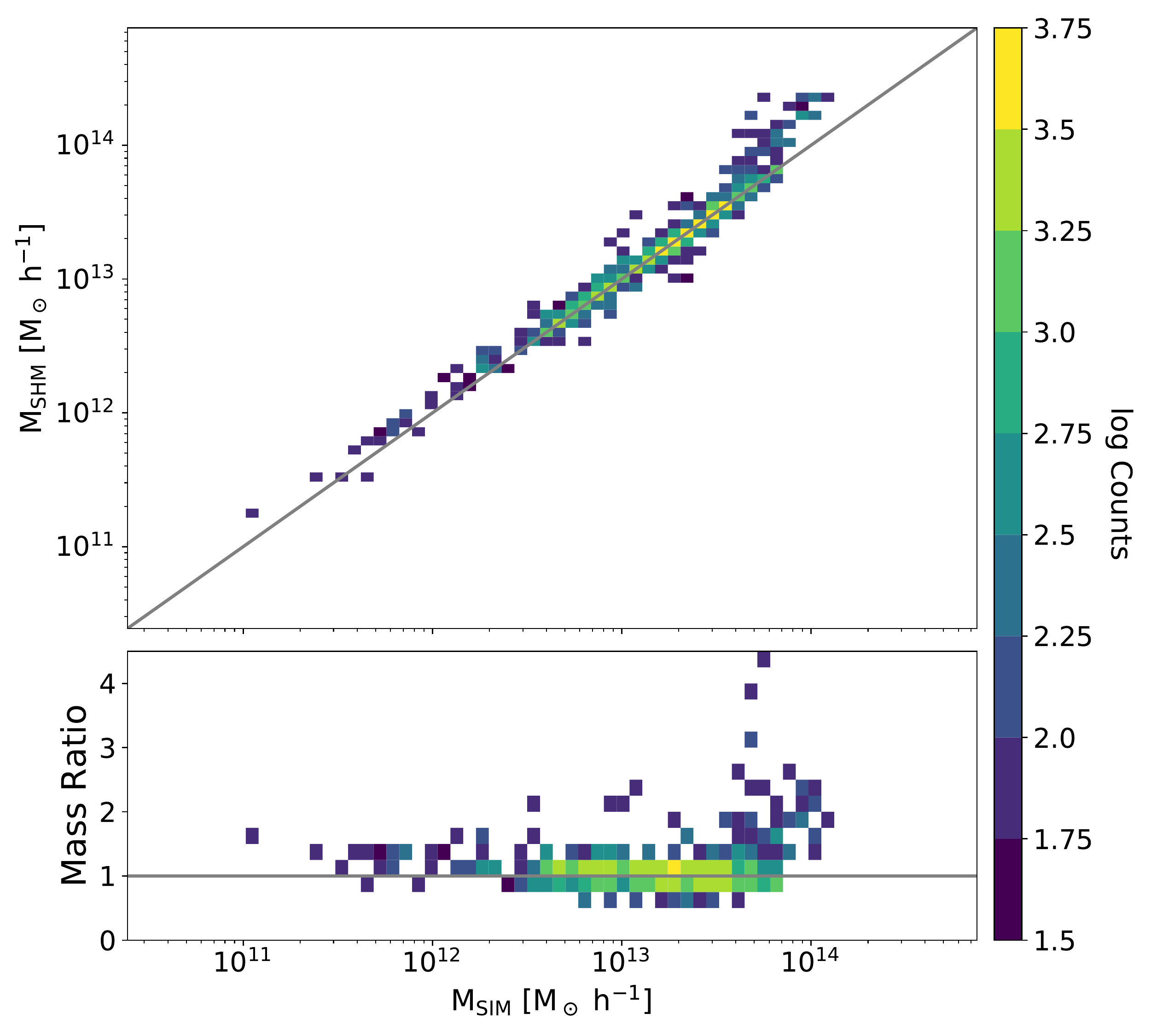}
\caption{\textsc{\textbf{Mass Comparison Between the $\MSHM$ and $\MSIM$.}} {\it Top panel}: Direct comparison between the aperture mass enclosed by the $\eER$ ($\MSHM$) and the ``true'' aperture mass within the same aperture from the simulation surface density ($\MSIM$). The solid gray line is where $\MSHM = \MSIM$, plotted to guide the eye. {\it Bottom panel}: The mass ratio, $\MSHM / \MSIM$. We find that on average, $\MSHM$ overestimates $\MSIM$, especially at the high mass bins (see \S\ref{sec:results}).}
\label{fig:MvM}
\end{figure}

In \autoref{fig:MvM}, we plot a direct comparison between $\MSHM$ and $\MSIM$. We measure an overall scatter of \AllScatter\ about the 1:1 relation between $\MSHM$ and $\MSIM$ (drawn in \autoref{fig:MvM} to guide the eye), with a positive bias of \AllBias. 
Interestingly, at the high-mass bin the core mass is highly overestimated; we explain this bias below.  

\begin{figure*}
\center
\includegraphics[width=1\textwidth]{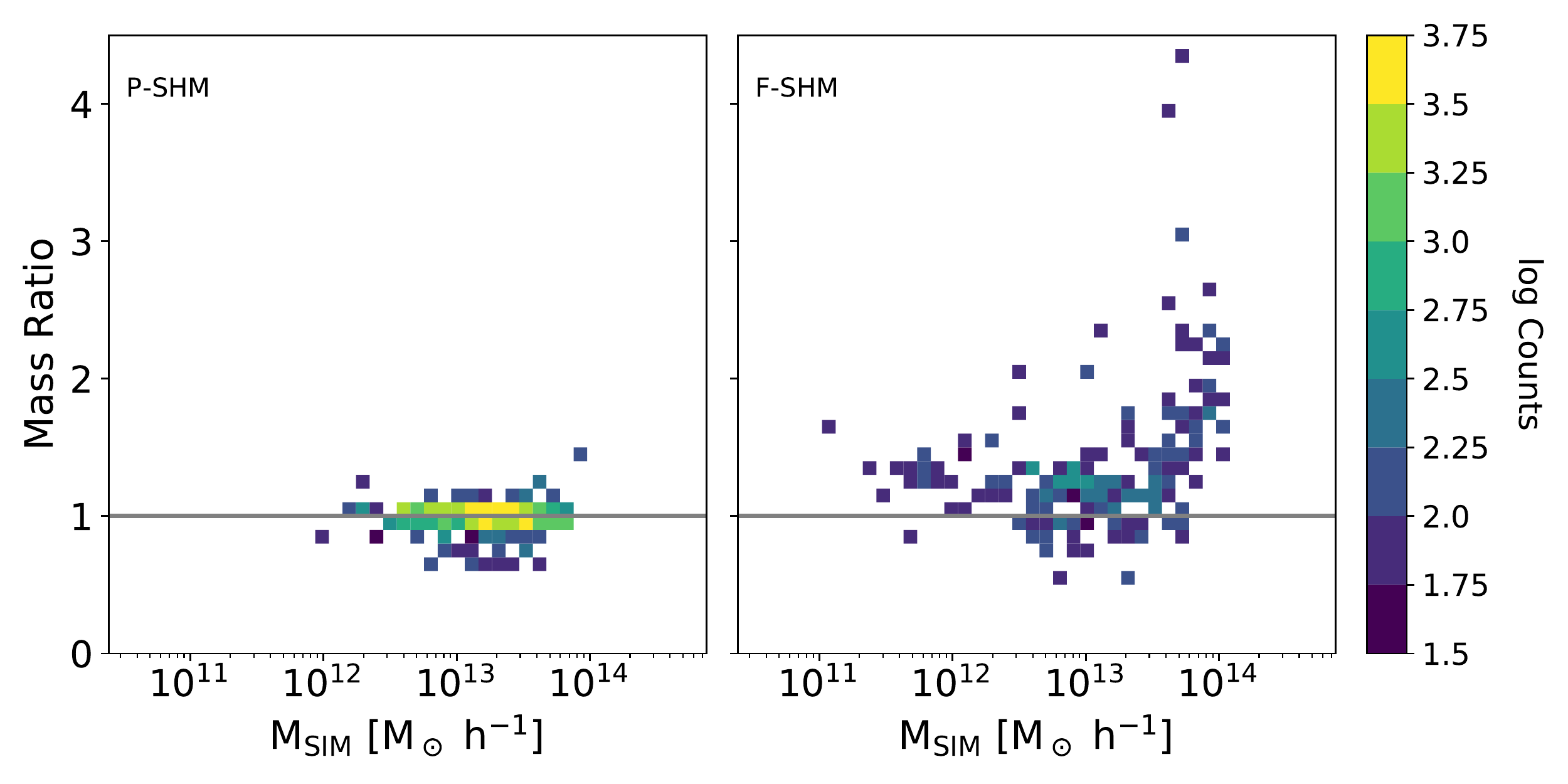}
\caption{\textsc{\textbf{Comparison between the mass estimates of P-SHMs and F-SHMs.}} Mass ratio ($\MSHM / \MSIM$) of the SHMs that passed the quick visual inspection (P-SHMs) is plotted in the left panel, and of the ones that Failed (F-SHMs) is plotted in the right panel. The P-SHMs span a somewhat smaller mass range than the F-SHMs. Notably, on average the F-SHMs are biased high, and their spread about the one-to-one line is is higher than the P-SHM, indicating that a quick visual inspection of the SHM outputs can easily weed out most of the outliers.}
\label{fig:g_v_ng_shm}
\end{figure*}

In \autoref{fig:g_v_ng_shm}, we separate the sample into two bins, according to their pass/fail assessment (see \S\ref{subsec:shm_visual_inspec}). 
We find that the P-SHMs span a tighter core-mass range compared to the F-SHMs, i.e., SHMs in the high and low mass bins are more likely to fail.
In all mass bins the core masses computed from the F-SHM models are less accurate; in particular, the large scatter observed in the high mass bin in \autoref{fig:MvM} is due entirely to F-SHM models. 
Overall, the mass estimate of the P-SHMs has a scatter of \GSHMScatter\ with a bias of \GSHMBias\ compared to the true mass. This result implies that the larger scatter and bias of the whole sample are driven by the failed-SHM lines of sight. 

Further investigation of these catastrophic failures highlights the limitation of SHM in recovering some under-constrained lensing configurations. We find that in most of the high-mass, highly overestimated cases, the SHM converges on a solution where the halo is oriented such that the single giant arc forms on a critical curve in the direction of its semi minor axis, rather than in the direction of its semi major axis. An example of such a failed model is shown in the bottom middle panel of \autoref{fig:ng_shm_ex}. These models populate the highest core-mass bin in \autoref{fig:MvM} since they produce a large $e\theta_E$; and they overestimate the mass since the model converges on a wrong solution, perpendicular to the underlying mass distribution. Fortunately, these cases are easily identified in a visual inspection. In analyses of real data, these cases can be flagged for a more involved analysis beyond the automated SHM. Manually imposing more constrained priors, increasing the complexity of the model, or adding constraints from secondary lensed system may resolve these cases.

All of these indicate that a quick visual inspection of the model outputs is beneficial when estimating the mass at the cores of galaxy clusters, using the SHM method.

\subsection{Possible causes of scatter and bias in the $\MSHM$ mass estimate}
\label{subsec:bin_analysis}

\begin{figure*}
\center
\includegraphics[width=1\textwidth]{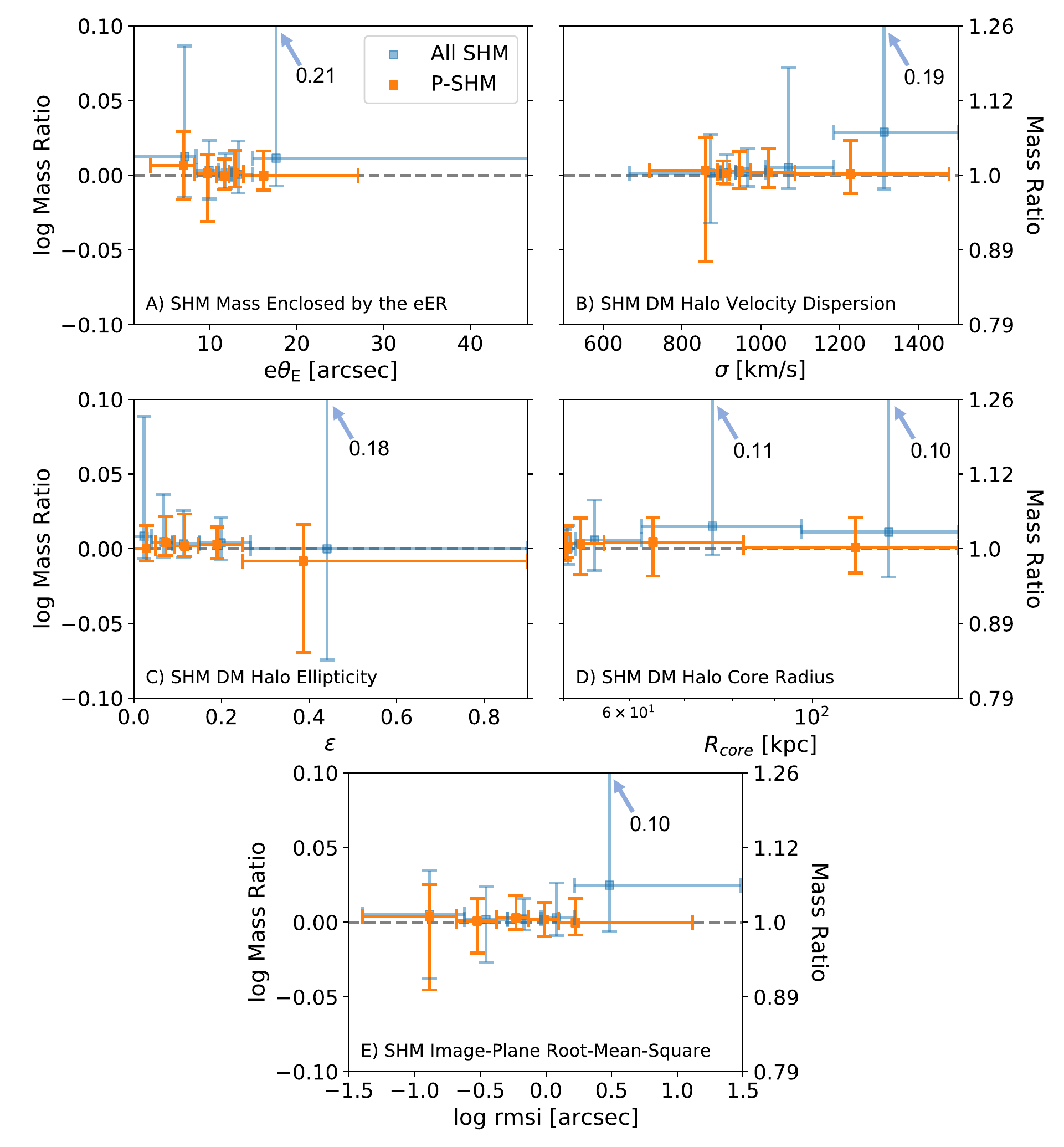}
\caption{\textsc{\textbf{Mass Ratio ($\MSHM / \MSIM$) Binned by the SHM Best Fit Parameters.}} Mass ratio, binned by the effective Einstein radius ($\eER$, panel A), dark matter (DM) halo model velocity dispersion ($\sigma$, panel B), DM halo model ellipticity ($\epsilon$, panel C), DM halo model core radius (R$_{core}$, panel D), and the image-plane root-mean-square (rmsi, panel E). The symbol marks the median of the distribution of the mass ratio, the horizontal error bars indicate the bin size (selected such that there is an equal number of SHMs per bin), and the vertical  error  bars  represent  the  16th  and  84th  percentile. We plot the results for all SHMs (blue) and only P-SHMs (orange) for comparison. 
We find that the P-SHMs have no bias, and a smaller scatter than un-inspected sample (all SHMs). Without eliminating the failed models, the scatter is higher overall; it increases with the extreme (low and high) values of model parameters. See \autoref{subsec:bin_analysis} for discussion.
}
\label{fig:shm_binned}
\end{figure*}

We explore possible dependencies in the scatter and bias of $\MSHM$ with respect to $\MSIM$ against the SHM best fit parameters (velocity dispersion ($\sigma$), ellipticity ($\epsilon$), and core radius ($R_{core}$)). The results are shown in \autoref{fig:shm_binned}. When considering the entire sample (including failed SHMs), we find that the scatter is larger at small and large $\epsilon$, $\eER$, and at large values of $R_{core}$.

For the models that pass the visual inspection, P-SHM (plotted in orange for comparison), we find no trends in the scatter or bias with any of the model parameters. This plot clearly shows the reduction in the scatter and bias in the P-SHMs when compared to all-SHMs. 

The large scatter and high occurrence of F-SHM found in the extreme values of the SHM fit parameters -- $\sigma$, $\epsilon$, and $R_{core}$ -- indicate that during the minimization process the best-fit model was found at the edge of the parameter space. These cases require additional human attention, better parameter exploration, and possibly an increase in the flexibility of the model. These interventions are not allowed in the framework of automated SMH, but are common practice in detailed lensing analyses. The increase in the complexity of the model is usually met with a need of additional constraints. It is expected that the SHM will struggle to reproduce dark matter halos with significantly disturbed morphology or mergers, and converge on, e.g., the highest ellipticity allowed, as shown in the bottom-left panel of \autoref{fig:ng_shm_ex}.

\begin{figure*}
\center
\includegraphics[width=1\textwidth]{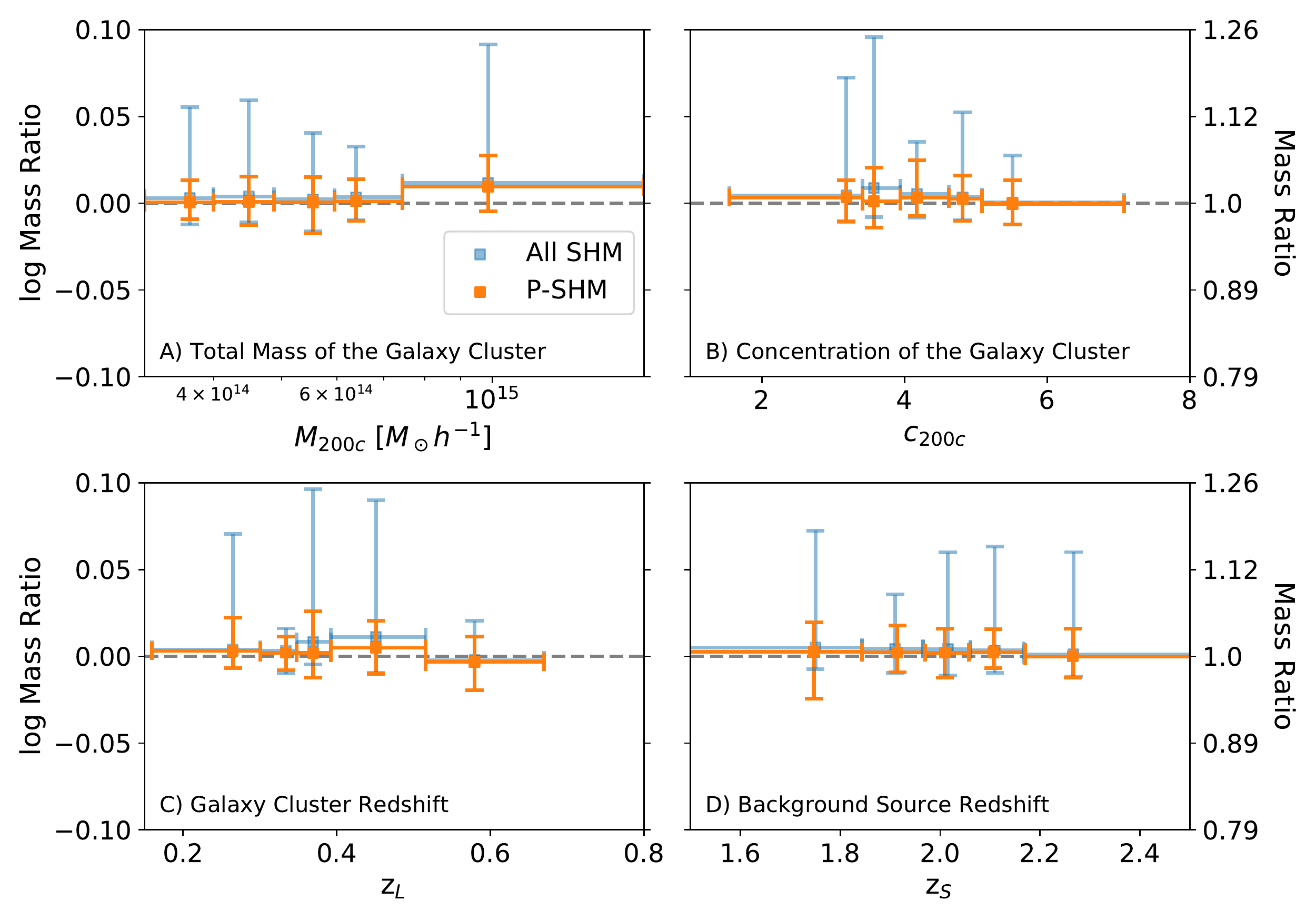}
\caption{\textsc{\textbf{Mass Ratio ($\MSHM / \MSIM$) Binned by the Lens-Background-Source System Properties.}} The mass ratio binned by the total mass ($\mathrm{M}_{200c}$, panel A), concentration ($c_{200c}$, panel B), galaxy cluster redshift ($\zl$, panel C), and background source redshift ($\zs$, panel D) are plotted for all SHMs (blue) and the P-SHMs (orange). The symbol marks the median of the distribution of the mass ratio, the horizontal error bars indicate the bin size (equal number of SHMs per bin), and the vertical  error  bars  represent  the  16th  and  84th  percentile. We find no trend in the scatter and bias with respect to any of the simulated cluster and background source properties for neither all SHMs nor P-SHMs.}
\label{fig:sim_binned}
\end{figure*}

We also test the results against the image-plane root-mean-square (\rmsi), in order to determine whether it could serve as a quantitative indicator of model quality (last panel of \autoref{fig:sim_binned}).
As expected, we find large scatter and bias in the high \rmsi\ bin, attributed to F-SHMs. However, we also find that both all-SHMs and the P-SHMs have high scatter in the lowest \rmsi\ bin (\rmsi$\approx 1\farcs0$). 
This behavior is consistent with previous studies. For example, \cite{Johnson:16} show that a larger number of constraints lead to a better accuracy in recovering the underlying mass and magnification, while the \rmsi\ becomes worse.
The trend of increased scatter with decreased \rmsi, and the fact that some F-SHM have low \rmsi, both indicate that the \rmsi\ does not provide a good indicator of model quality. However, high \rmsi\ values may be useful as an initial triage to remove some of the catastrophic failures before visual inspection.

For reference, if excluding models with $\rm{log}$(\rmsi) $> 0.5$, the scatter and bias reduce to $4.8\%$ and $0.65\%$, respectively, better than the overall sample (\AllScatter, \AllBias), but not as good as the P-SHM sample (\GSHMScatter, \GSHMBias).

Next, we explore whether the scatter and bias depend on properties of the simulated galaxy cluster -- total mass ($\mathrm{M}_{200c}$), concentration (c$_{200c}$), cluster redshift ($\zl$), and background sources redshift ($\zs$). The results are shown in \autoref{fig:sim_binned}. We find a flat trend in the scatter and bias with respect to all of the cluster and background source properties for both the SHMs (including F-SHMs) and P-SHMs. This exploration of the scatter and bias is crucial for future studies, that may use the $\MSHM$ method to measure the core mass and combine it with a large scale mass proxy to measure, e.g., the concentration of an ensemble of galaxy clusters. Based on this result, we conclude that using $\MSHM$ to measure core masses will not bias such future work.

\section{The Effect of the Background Source Redshift and the Lensing Configuration of the Arcs}
\label{sec:no_zs_and_lens_config}

In this section, we investigate the effect on the Single-Halo model and the aperture mass enclosed by the effective Einstein radius of the SHM due to the background source redshift, lensing configuration of the arcs, and addition of a second lensed image system. Here we do not exclude the F-SHMs from the analysis.

\subsection{Effects of the Background Source Redshift ($\zs$) on $\MSHM$}
\label{subsec:no_zs}

When a secure spectroscopic redshift of a lensed galaxy  is not available, lens modelers often leave the source redshift as a free parameter, sometimes using its photometric redshift as prior. By leaving the background source redshift as a free parameter in our test models, the number of degrees of freedom increases to seven, requiring seven or more constraints. This is satisfied by $895$ ray-traced images in our overall simulated sample. We apply a broad uniform prior on the background source redshift, $1 \leq \zs \leq 5$. From the computed models, we find that the model-predicted redshifts are on average $1.9$ times higher than the true redshifts. We measure a scatter of \NOZSSHMScatter\ with a bias of \NOZSSHMBias\ on the mass estimate $\MSHM$. The $\MSHM$, when no background source redshift is known, underestimates the true aperture mass enclosed by the effective Einstein radius and the scatter increases. This effect shows the degeneracy between the derived mass and the background source redshift, and highlights the importance of securing spectroscopic redshifts of background sources for the accuracy of lens models. 

\subsection{Effects of the Lensing Configuration on $\MSHM$}
\label{subsec:lens_config}

\begin{figure*}
\center
\includegraphics[width=1\textwidth]{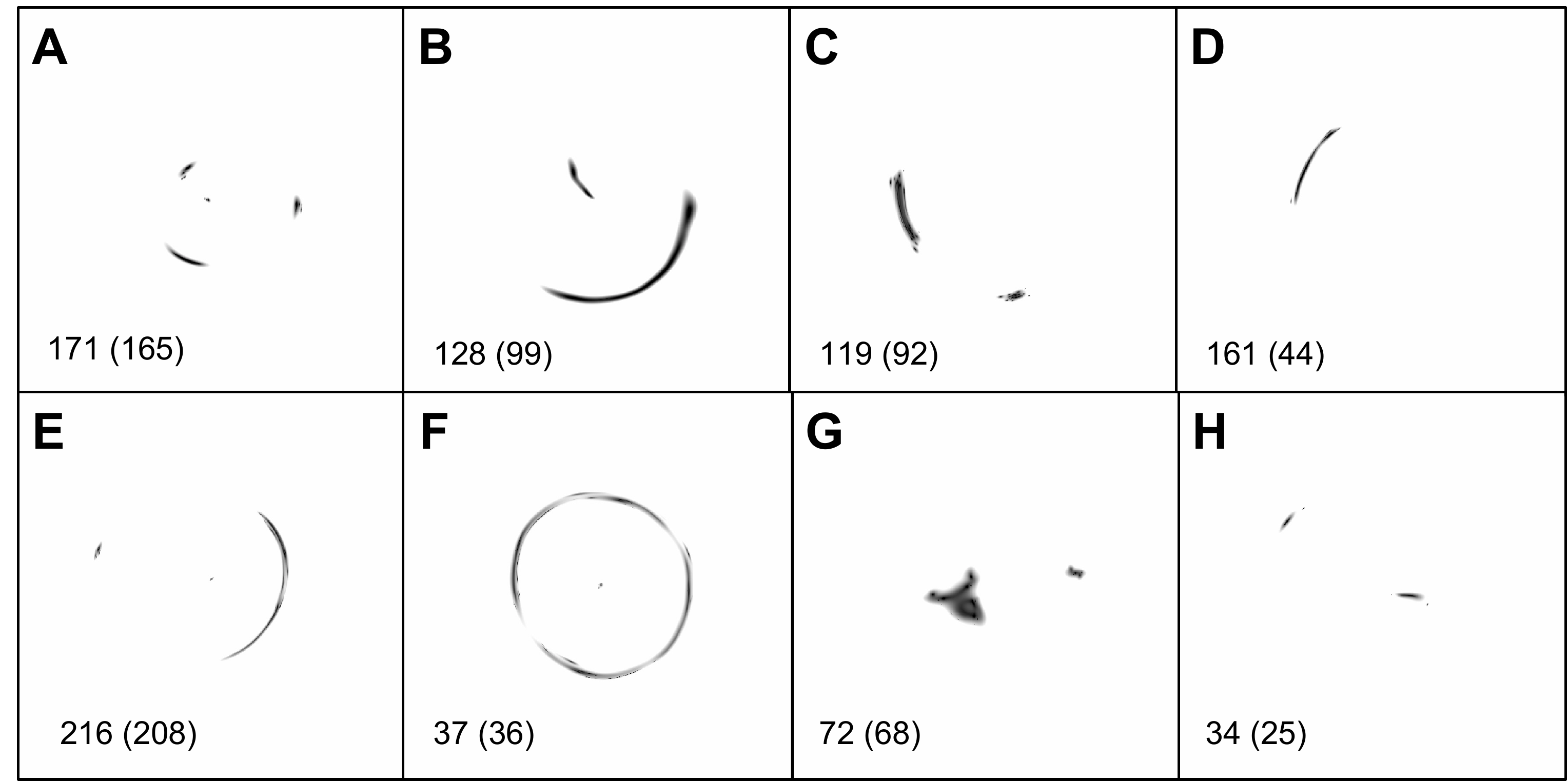}
\caption{\textsc{\textbf{Strong Lensing Geometric Configurations of Arcs.}} Examples of the groups of the simulated SL geometric configurations of the arcs. Each image is $1.0$ arcminute on the side. In the bottom left corner of every panel we indicate the total number of simulated images in that group and in parenthesis the number of SHMs that pass visual inspection. See \autoref{subsec:lens_config} for a description of the different groups.}
\label{fig:sl_config}
\end{figure*}

We explore the effect of the lensing configuration of the simulated images on the accuracy of the SHMs, as different configurations provide different constraining power. We inspect each of the simulated lensed images and sort them into eight groups of similar lensing geometry. A representative example of each of the groups is shown in \autoref{fig:sl_config}, along with the number of simulated images in each group. Systems in Group A typically have five images: a merging pair forming a tangential arc, a counter image, and a clearly observable pair of radial arcs. Systems in Group B show a single merging tangential arc and a clearly observable pair of radial arcs. Systems in Group C have a similar configuration to group A,  without visible radial arcs. Systems in Group D have a single arc similar to group B, but without visible radial arcs. Systems in Group E, have a tangential arc made of a merging pair and a counter image, but unlike group A only a single radial arc is identified. Group F includes the Einstein ring configuration. Systems in Group G have a set of radial and tangential arcs close to each other and an additional counter image. Group H systems form a merging pair of radial arcs and a single tangential arc.

The SL configuration group that has the most F-SHMs is group D (a single tangential arc);  out of the $161$ lensed images with this configuration, $117$ ($\sim 73\%$) are F-SHMs and $44$ ($\sim 27\%$) are P-SHMs. This group accounts for more than half of the $201$ total F-SHMs.
This lensing configuration, of a single giant arc, provides the least geometrical constraining power, as it leaves regions of the lens plane with no constraints. 
Since the model is only locally constrained, lensing configurations in which the single halo is oriented approximately perpendicular to the orientation of the underlying mass distribution are allowed; models that have constrains only on one side of the center of mass suffer from high degeneracy between the halo position, its ellipticity, and velocity dispersion. Such low constraining power is also reported in some observed systems with single giant arcs in \cite{Sharon:20}. The low constraining power can therefore result in unphysical SHMs and an unreliable measurement of the mass enclosed by the effective Einstein radius.
Excluding the $161$ single-arc images from the full sample, the scatter and bias reduce to \SLCONFSHMScatter\ and \SLCONFSHMBias\,respectively, significantly improving upon the scatter of the full sample (scatter: \AllScatter, bias: \AllBias), and close to the precision of the P-SHM only sample (scatter: \GSHMScatter, bias: \GSHMBias). 

Some of the challenges inherent to using a single giant arc as the only constraint can be mitigated by obtaining more lensing constraints (see \S\ref{subsec:second_arc}), which often requires higher resolution or deeper imaging. Some of these failed cases can be recovered with a more involved analysis, manual inspection of the parameter space, inclusion of physically-motivated priors (such as an upper-limit on the velocity dispersion based cluster richness). As noted in \S\ref{subsec:bin_analysis}, these interventions are beyond the framework of the quasi-automated SHM.

\subsection{Constraining Power of Secondary Lensed Image Systems}
\label{subsec:second_arc}

Last, we investigate the constraining power of a secondary lensed source system. The SHM benefits from the addition of a lensed system when these new constraints complement the geometrical constraints of the primary image system and the mass distribution. The increase in constraining power at the core of the galaxy cluster benefits lensed systems with a single giant arc (Group D) where the arc coverage is limited to one side of the cluster.

For example, when adding a secondary lensed system (of any non-D group) to failed lens models of group D, $53\%$ of these formerly F-SHM pass the visual inspection and become P-SHM. This behavior illustrates the importance of identifying additional lensed images to help constrain the lens models.
On the other hand, about a third of the failed D models did not benefit from adding a secondary system, even though the SHM produced a passing model when using only that second system as constraints. A further inspection of these failed models indicates that in these particular cases, the lensing geometry of the D arc was so hard for a single-halo model to reproduce that it forced the model to converge on a failed solution. Such configurations will likely require more involved lensing analysis, possibly with higher flexibility in the lens model or imposing user-identified priors beyond the automated SHM process.


\section{Conclusions} 
\label{sec:conclusion}

We explore the use of Single-Halo models as an automated tool to efficiently estimate the mass at the core of SL galaxy clusters. The SHM can be automatically computed once the arcs have been identified and background source redshift measured. This method uses the parametric lensing algorithm \lenstool\ with a single dark matter halo, represented by a dPIE. The constraints used are the lens redshift, the positions of the lensed images, and their source redshift. An initial halo center position (e.g. the BCG) is also needed as an input. To characterize the scatter and bias in the estimator, we use ray-traced simulated images from the state-of-the-art Outer Rim simulation. We compute the SHM, measure the aperture mass enclosed by the effective Einstein radius ($\MSHM$), and compare the mass estimate to the mass from the simulation, measured within the same aperture, ($\MSIM$). We conclude the following:

\begin{itemize}
	\item Considering the entire sample, the scatter of $\MSHM$ is \AllScatter\ with a bias of \AllBias\ compared to the true mass, $\MSIM$.
	\item A quick visual inspection of the single halo models reveals that some fail to reproduce the lensing configuration or lead to unphysical lens models. Excluding the failed SHMs (F-SHM) reduces the scatter and bias to \GSHMScatter\ and \GSHMBias, respectively.
	\item We find that the scatter is larger at small and large values of the ellipticity ($\epsilon$) and effective Einstein radius ($\eER$), and large values of the core radius ($R_{core}$). Excluding the F-SHMs eliminates this trend.
	\item We find a weak increasing trend in the bias with the SHM velocity dispersion ($\sigma$), and a larger scatter at larger $\sigma$. Excluding the F-SHMs eliminates this trend.
	\item We find no significant dependence on the bias and scatter of $\MSHM$ with respect to the properties of the lens-source system -- total mass, concentration, lens redshift, and source redshift. This exploration is crucial for future studies, that may use the $\MSHM$ method to measure the core mass and combine it with a large scale mass proxy to measure, e.g., the concentration of an ensemble of galaxy clusters. Based on this result, we conclude that using $\MSHM$ to measure core masses will not bias such future work.
	\item A high \rmsi\ can be used to identify and eliminate some of the worst cases of F-SHM, before the visual inspection; but on its own, it is an insufficient predictor of the SHM model quality.
	\item When the background source redshift is unknown, lens modelers frequently use priors on the source redshift and set the redshift as a free parameter in the lens model. We use a broad uniform prior on the source redshift and find that the model-predicted redshifts are overestimated, leading to an underestimate of the mass. When the source redshift is unknown the scatter and bias $\MSHM$ are \NOZSSHMScatter\ and \NOZSSHMBias\ compared to $\MSIM$ respectively. 
	Our analysis affirms the importance of securing spectroscopic redshifts or high-quality photometric redshifts for the lensing constraints, in order to obtain a precise and accurate mass measurement. This is consistent with findings by e.g., \citet{Caminha:16,Johnson:16}.
	\item The lensing configuration affects the efficacy of the SHM. We find that a single arc configuration (group D) provides the least constraining power and accounts for most of the extreme outliers. Excluding these systems, we compute a scatter and bias on $\MSHM$ of \SLCONFSHMScatter\ and \SLCONFSHMBias\ against $\MSIM$, respectively.
	\item The addition of a second lensed source helps constrain the lens model and particularly benefits lens models where the geometrical configuration of the arcs has limited constraining power at the core of the SL galaxy clusters (group D). It is therefore most cost-effective to follow-up systems with a single arc with deeper or high resolution imaging, in order to secure additional lensing constraints. However, some lensing configurations may require a more complex lens modeling process than the SHM.
\end{itemize}

In the future, new tools will expedite the current manual process of strong lensing analysis. Examples include the introduction of convolutional neural networks for identification of strong lensing features (e.g., \citealt{Petrillo:17,Jacobs:19,Canameras:20,HuangX:20}), and machine learning algorithms to model the mass distribution of strong lenses (e.g., \citealt{Bom:19,Pearson:19}). We look forward to the continuous development of these tools as the Single-Halo models introduced in this work will greatly benefit from them.

The evaluation of $\MSHM$ presented in this work, facilitates the use of this efficient mass estimate at the core of SL galaxy clusters, and enables an automated measurement of the core mass in the large samples of strong lensing clusters from current and future surveys.


\section*{Acknowledgements}

The authors would like to thank the anonymous referee for insightful suggestions that improved this manuscript. This material is based upon work supported by the National Science Foundation Graduate Research Fellowship Program under Grant No. DGE 1256260. Work at Argonne National  Lab is supported by UChicago Argonne LLC, Operator of Argonne National Laboratory. Argonne National Lab, a U.S. Department of Energy Office of Science Laboratory, is operated by UChicago Argonne LLC under contract no. DE-AC02-06CH11357. This research used resources of the Argonne Leadership Computing Facility, which is a DOE Office of Science User Facility supported under Contract DE-AC02-06CH11357.


\bibliographystyle{yahapj.bst}
\bibliography{bibfile.bib}

\begin{thebibliography}{}
\providecommand\natexlab[1]{#1}
\providecommand\JournalTitle[1]{#1}

\bibitem[{{Acebron} {et~al.}(2017){Acebron}, {Jullo}, {Limousin}, {Tilquin},
  {Giocoli}, {Jauzac}, {Mahler}, \& {Richard}}]{Acebron:17}
{Acebron}, A., {Jullo}, E., {Limousin}, M., {et~al.} 2017,
  \href{http://dx.doi.org/10.1093/mnras/stx1330}{\JournalTitle{\mnras}, 470,
  1809}

\bibitem[{{Allen} {et~al.}(2011){Allen}, {Evrard}, \& {Mantz}}]{Allen:11}
{Allen}, S.~W., {Evrard}, A.~E., \& {Mantz}, A.~B. 2011,
  \href{http://dx.doi.org/10.1146/annurev-astro-081710-102514}{\JournalTitle{Annual
  Review of Astronomy and Astrophysics}, 49, 409}

\bibitem[{{Amendola} {et~al.}(2018){Amendola}, {Appleby}, {Avgoustidis},
  {Bacon}, {Baker}, {Baldi}, {Bartolo}, {Blanchard}, {Bonvin}, {Borgani},
  {Branchini}, {Burrage}, {Camera}, {Carbone}, {Casarini}, {Cropper}, {de
  Rham}, {Dietrich}, {Di Porto}, {Durrer}, {Ealet}, {Ferreira}, {Finelli},
  {Garc{\'\i}a-Bellido}, {Giannantonio}, {Guzzo}, {Heavens}, {Heisenberg},
  {Heymans}, {Hoekstra}, {Hollenstein}, {Holmes}, {Hwang}, {Jahnke},
  {Kitching}, {Koivisto}, {Kunz}, {La Vacca}, {Linder}, {March}, {Marra},
  {Martins}, {Majerotto}, {Markovic}, {Marsh}, {Marulli}, {Massey}, {Mellier},
  {Montanari}, {Mota}, {Nunes}, {Percival}, {Pettorino}, {Porciani},
  {Quercellini}, {Read}, {Rinaldi}, {Sapone}, {Sawicki}, {Scaramella},
  {Skordis}, {Simpson}, {Taylor}, {Thomas}, {Trotta}, {Verde}, {Vernizzi},
  {Vollmer}, {Wang}, {Weller}, \& {Zlosnik}}]{Amendola:18}
{Amendola}, L., {Appleby}, S., {Avgoustidis}, A., {et~al.} 2018,
  \href{http://dx.doi.org/10.1007/s41114-017-0010-3}{\JournalTitle{Living
  Reviews in Relativity}, 21, 2}

\bibitem[{{Bayliss} {et~al.}(2011){Bayliss}, {Gladders}, {Oguri}, {Hennawi},
  {Sharon}, {Koester}, \& {Dahle}}]{Bayliss:11}
{Bayliss}, M.~B., {Gladders}, M.~D., {Oguri}, M., {et~al.} 2011,
  \href{http://dx.doi.org/10.1088/2041-8205/727/1/L26}{\JournalTitle{\apj},
  727, L26}

\bibitem[{{Bayliss} {et~al.}(2014){Bayliss}, {Johnson}, {Gladders}, {Sharon},
  \& {Oguri}}]{Bayliss:14}
{Bayliss}, M.~B., {Johnson}, T., {Gladders}, M.~D., {Sharon}, K., \& {Oguri},
  M. 2014,
  \href{http://dx.doi.org/10.1088/0004-637X/783/1/41}{\JournalTitle{\apj}, 783,
  41}

\bibitem[{{Benson} {et~al.}(2014){Benson}, {Ade}, {Ahmed}, {Allen}, {Arnold},
  {Austermann}, {Bender}, {Bleem}, {Carlstrom}, {Chang}, {Cho}, {Cliche},
  {Crawford}, {Cukierman}, {de Haan}, {Dobbs}, {Dutcher}, {Everett}, {Gilbert},
  {Halverson}, {Hanson}, {Harrington}, {Hattori}, {Henning}, {Hilton},
  {Holder}, {Holzapfel}, {Irwin}, {Keisler}, {Knox}, {Kubik}, {Kuo}, {Lee},
  {Leitch}, {Li}, {McDonald}, {Meyer}, {Montgomery}, {Myers}, {Natoli},
  {Nguyen}, {Novosad}, {Padin}, {Pan}, {Pearson}, {Reichardt}, {Ruhl},
  {Saliwanchik}, {Simard}, {Smecher}, {Sayre}, {Shirokoff}, {Stark}, {Story},
  {Suzuki}, {Thompson}, {Tucker}, {Vanderlinde}, {Vieira}, {Vikhlinin}, {Wang},
  {Yefremenko}, \& {Yoon}}]{Benson:14}
{Benson}, B.~A., {Ade}, P.~A.~R., {Ahmed}, Z., {et~al.} 2014,
  \href{http://dx.doi.org/10.1117/12.2057305}{in Millimeter, Submillimeter, and
  Far-Infrared Detectors and Instrumentation for Astronomy VII, Vol. 9153},
  91531P

\bibitem[{{Bleem} {et~al.}(2015){Bleem}, {Stalder}, {de Haan}, {Aird}, {Allen},
  {Applegate}, {Ashby}, {Bautz}, {Bayliss}, {Benson}, {Bocquet}, {Brodwin},
  {Carlstrom}, {Chang}, {Chiu}, {Cho}, {Clocchiatti}, {Crawford}, {Crites},
  {Desai}, {Dietrich}, {Dobbs}, {Foley}, {Forman}, {George}, {Gladders},
  {Gonzalez}, {Halverson}, {Hennig}, {Hoekstra}, {Holder}, {Holzapfel},
  {Hrubes}, {Jones}, {Keisler}, {Knox}, {Lee}, {Leitch}, {Liu}, {Lueker},
  {Luong-Van}, {Mantz}, {Marrone}, {McDonald}, {McMahon}, {Meyer}, {Mocanu},
  {Mohr}, {Murray}, {Padin}, {Pryke}, {Reichardt}, {Rest}, {Ruel}, {Ruhl},
  {Saliwanchik}, {Saro}, {Sayre}, {Schaffer}, {Schrabback}, {Shirokoff},
  {Song}, {Spieler}, {Stanford}, {Staniszewski}, {Stark}, {Story}, {Stubbs},
  {Vanderlinde}, {Vieira}, {Vikhlinin}, {Williamson}, {Zahn}, \&
  {Zenteno}}]{Bleem:15}
{Bleem}, L.~E., {Stalder}, B., {de Haan}, T., {et~al.} 2015,
  \href{http://dx.doi.org/10.1088/0067-0049/216/2/27}{\JournalTitle{The
  Astrophysical Journal Supplement Series}, 216, 27}

\bibitem[{{Bleem} {et~al.}(2020){Bleem}, {Bocquet}, {Stalder}, {Gladders},
  {Ade}, {Allen}, {Anderson}, {Annis}, {Ashby}, {Austermann}, {Avila}, {Avva},
  {Bayliss}, {Beall}, {Bechtol}, {Bender}, {Benson}, {Bertin}, {Bianchini},
  {Blake}, {Brodwin}, {Brooks}, {Buckley-Geer}, {Burke}, {Carlstrom}, {Rosell},
  {Carrasco Kind}, {Carretero}, {Chang}, {Chiang}, {Citron}, {Moran},
  {Costanzi}, {Crawford}, {Crites}, {da Costa}, {de Haan}, {De Vicente},
  {Desai}, {Diehl}, {Dietrich}, {Dobbs}, {Eifler}, {Everett}, {Flaugher},
  {Floyd}, {Frieman}, {Gallicchio}, {Garc{\'\i}a-Bellido}, {George}, {Gerdes},
  {Gilbert}, {Gruen}, {Gruendl}, {Gschwend}, {Gupta}, {Gutierrez}, {Halverson},
  {Harrington}, {Henning}, {Heymans}, {Holder}, {Hollowood}, {Holzapfel},
  {Honscheid}, {Hrubes}, {Huang}, {Hubmayr}, {Irwin}, {James}, {Jeltema},
  {Joudaki}, {Khullar}, {Klein}, {Knox}, {Kuropatkin}, {Lee}, {Li}, {Lidman},
  {Lowitz}, {MacCrann}, {Mahler}, {Maia}, {Marshall}, {McDonald}, {McMahon},
  {Melchior}, {Menanteau}, {Meyer}, {Miquel}, {Mocanu}, {Mohr}, {Montgomery},
  {Nadolski}, {Natoli}, {Nibarger}, {Noble}, {Novosad}, {Padin}, {Palmese},
  {Parkinson}, {Patil}, {Paz-Chinch{\'o}n}, {Plazas}, {Pryke}, {Ramachandra},
  {Reichardt}, {Remolina Gonz{\'a}lez}, {Romer}, {Roodman}, {Ruhl}, {Rykoff},
  {Saliwanchik}, {Sanchez}, {Saro}, {Sayre}, {Schaffer}, {Schrabback},
  {Serrano}, {Sharon}, {Sievers}, {Smecher}, {Smith}, {Soares-Santos}, {Stark},
  {Story}, {Suchyta}, {Tarle}, {Tucker}, {Vanderlinde}, {Veach}, {Vieira},
  {Wang}, {Weller}, {Whitehorn}, {Wu}, {Yefremenko}, \& {Zhang}}]{Bleem:20}
{Bleem}, L.~E., {Bocquet}, S., {Stalder}, B., {et~al.} 2020,
  \href{http://dx.doi.org/10.3847/1538-4365/ab6993}{\JournalTitle{\apjs}, 247,
  25}

\bibitem[{{Bocquet} {et~al.}(2020){Bocquet}, {Heitmann}, {Habib}, {Lawrence},
  {Uram}, {Frontiere}, {Pope}, \& {Finkel}}]{Bocquet:20}
{Bocquet}, S., {Heitmann}, K., {Habib}, S., {et~al.} 2020,
  \href{http://dx.doi.org/10.3847/1538-4357/abac5c}{\JournalTitle{\apj}, 901,
  5}

\bibitem[{{Bocquet} {et~al.}(2019){Bocquet}, {Dietrich}, {Schrabback}, {Bleem},
  {Klein}, {Allen}, {Applegate}, {Ashby}, {Bautz}, {Bayliss}, {Benson},
  {Brodwin}, {Bulbul}, {Canning}, {Capasso}, {Carlstrom}, {Chang}, {Chiu},
  {Cho}, {Clocchiatti}, {Crawford}, {Crites}, {de Haan}, {Desai}, {Dobbs},
  {Foley}, {Forman}, {Garmire}, {George}, {Gladders}, {Gonzalez}, {Grandis},
  {Gupta}, {Halverson}, {Hlavacek-Larrondo}, {Hoekstra}, {Holder}, {Holzapfel},
  {Hou}, {Hrubes}, {Huang}, {Jones}, {Khullar}, {Knox}, {Kraft}, {Lee}, {von
  der Linden}, {Luong-Van}, {Mantz}, {Marrone}, {McDonald}, {McMahon}, {Meyer},
  {Mocanu}, {Mohr}, {Morris}, {Padin}, {Patil}, {Pryke}, {Rapetti},
  {Reichardt}, {Rest}, {Ruhl}, {Saliwanchik}, {Saro}, {Sayre}, {Schaffer},
  {Shirokoff}, {Stalder}, {Stanford}, {Staniszewski}, {Stark}, {Story},
  {Strazzullo}, {Stubbs}, {Vanderlinde}, {Vieira}, {Vikhlinin}, {Williamson},
  \& {Zenteno}}]{Bocquet:19}
{Bocquet}, S., {Dietrich}, J.~P., {Schrabback}, T., {et~al.} 2019,
  \href{http://dx.doi.org/10.3847/1538-4357/ab1f10}{\JournalTitle{\apj}, 878,
  55}

\bibitem[{{Bom} {et~al.}(2019){Bom}, {Poh}, {Nord}, {Blanco-Valentin}, \&
  {Dias}}]{Bom:19}
{Bom}, C., {Poh}, J., {Nord}, B., {Blanco-Valentin}, M., \& {Dias}, L. 2019,
  \JournalTitle{arXiv e-prints}, arXiv:1911.06341

\bibitem[{{Broadhurst} \& {Barkana}(2008)}]{Broadhurst:08}
{Broadhurst}, T.~J., \& {Barkana}, R. 2008,
  \href{http://dx.doi.org/10.1111/j.1365-2966.2008.13852.x}{\JournalTitle{\mnras},
  390, 1647}

\bibitem[{{Ca{\~n}ameras} {et~al.}(2020){Ca{\~n}ameras}, {Schuldt}, {Suyu},
  {Taubenberger}, {Meinhardt}, {Leal-Taix{\'e}}, {Lemon}, {Rojas}, \&
  {Savary}}]{Canameras:20}
{Ca{\~n}ameras}, R., {Schuldt}, S., {Suyu}, S.~H., {et~al.} 2020,
  \href{http://dx.doi.org/10.1051/0004-6361/202038219}{\JournalTitle{\aap},
  644, A163}

\bibitem[{{Caminha} {et~al.}(2016){Caminha}, {Grillo}, {Rosati}, {Balestra},
  {Karman}, {Lombardi}, {Mercurio}, {Nonino}, {Tozzi}, {Zitrin}, {Biviano},
  {Girardi}, {Koekemoer}, {Melchior}, {Meneghetti}, {Munari}, {Suyu}, {Umetsu},
  {Annunziatella}, {Borgani}, {Broadhurst}, {Caputi}, {Coe}, {Delgado-Correal},
  {Ettori}, {Fritz}, {Frye}, {Gobat}, {Maier}, {Monna}, {Postman}, {Sartoris},
  {Seitz}, {Vanzella}, \& {Ziegler}}]{Caminha:16}
{Caminha}, G.~B., {Grillo}, C., {Rosati}, P., {et~al.} 2016,
  \href{http://dx.doi.org/10.1051/0004-6361/201527670}{\JournalTitle{\aap},
  587, A80}

\bibitem[{{Cerny} {et~al.}(2018){Cerny}, {Sharon}, {Andrade-Santos}, {Avila},
  {Brada{\v{c}}}, {Bradley}, {Carrasco}, {Coe}, {Czakon}, {Dawson}, {Frye},
  {Hoag}, {Huang}, {Johnson}, {Jones}, {Lam}, {Lovisari}, {Mainali}, {Oesch},
  {Ogaz}, {Past}, {Paterno-Mahler}, {Peterson}, {Riess}, {Rodney}, {Ryan},
  {Salmon}, {Sendra-Server}, {Stark}, {Strolger}, {Trenti}, {Umetsu},
  {Vulcani}, \& {Zitrin}}]{Cerny:18}
{Cerny}, C., {Sharon}, K., {Andrade-Santos}, F., {et~al.} 2018,
  \href{http://dx.doi.org/10.3847/1538-4357/aabe7b}{\JournalTitle{\apj}, 859,
  159}

\bibitem[{{Child} {et~al.}(2018){Child}, {Habib}, {Heitmann}, {Frontiere},
  {Finkel}, {Pope}, \& {Morozov}}]{Child:18}
{Child}, H.~L., {Habib}, S., {Heitmann}, K., {et~al.} 2018,
  \href{http://dx.doi.org/10.3847/1538-4357/aabf95}{\JournalTitle{\apj}, 859,
  55}

\bibitem[{{Chiriv{\`\i}} {et~al.}(2018){Chiriv{\`\i}}, {Suyu}, {Grillo},
  {Halkola}, {Balestra}, {Caminha}, {Mercurio}, \& {Rosati}}]{Chiviri:18}
{Chiriv{\`\i}}, G., {Suyu}, S.~H., {Grillo}, C., {et~al.} 2018,
  \href{http://dx.doi.org/10.1051/0004-6361/201731433}{\JournalTitle{\aap},
  614, A8}

\bibitem[{{Corless} \& {King}(2009)}]{Corless:09}
{Corless}, V.~L., \& {King}, L.~J. 2009,
  \href{http://dx.doi.org/10.1111/j.1365-2966.2009.14542.x}{\JournalTitle{\mnras},
  396, 315}

\bibitem[{{D'Aloisio} {et~al.}(2014){D'Aloisio}, {Natarajan}, \&
  {Shapiro}}]{DAloisio:14}
{D'Aloisio}, A., {Natarajan}, P., \& {Shapiro}, P.~R. 2014,
  \href{http://dx.doi.org/10.1093/mnras/stu1931}{\JournalTitle{\mnras}, 445,
  3581}

\bibitem[{{Diego} {et~al.}(2016){Diego}, {Broadhurst}, {Wong}, {Silk}, {Lim},
  {Zheng}, {Lam}, \& {Ford}}]{Diego:16}
{Diego}, J.~M., {Broadhurst}, T., {Wong}, J., {et~al.} 2016,
  \href{http://dx.doi.org/10.1093/mnras/stw865}{\JournalTitle{\mnras}, 459,
  3447}

\bibitem[{{Duffy} {et~al.}(2008){Duffy}, {Schaye}, {Kay}, \& {Dalla
  Vecchia}}]{Duffy:08}
{Duffy}, A.~R., {Schaye}, J., {Kay}, S.~T., \& {Dalla Vecchia}, C. 2008,
  \href{http://dx.doi.org/10.1111/j.1745-3933.2008.00537.x}{\JournalTitle{\mnras},
  390, L64}

\bibitem[{{El{\'\i}asd{\'o}ttir} {et~al.}(2007){El{\'\i}asd{\'o}ttir},
  {Limousin}, {Richard}, {Hjorth}, {Kneib}, {Natarajan}, {Pedersen}, {Jullo},
  \& {Paraficz}}]{Eliasdottir:07}
{El{\'\i}asd{\'o}ttir}, {\'A}., {Limousin}, M., {Richard}, J., {et~al.} 2007,
  \JournalTitle{arXiv e-prints}, arXiv:0710.5636

\bibitem[{{Evrard} {et~al.}(2002){Evrard}, {MacFarland}, {Couchman}, {Colberg},
  {Yoshida}, {White}, {Jenkins}, {Frenk}, {Pearce}, {Peacock}, \&
  {Thomas}}]{Evrard:02}
{Evrard}, A.~E., {MacFarland}, T.~J., {Couchman}, H.~M.~P., {et~al.} 2002,
  \href{http://dx.doi.org/10.1086/340551}{\JournalTitle{\apj}, 573, 7}

\bibitem[{{Gonzalez} {et~al.}(2012){Gonzalez}, {Stanford}, {Brodwin}, {Fedeli},
  {Dey}, {Eisenhardt}, {Mancone}, {Stern}, \& {Zeimann}}]{Gonzalez:12}
{Gonzalez}, A.~H., {Stanford}, S.~A., {Brodwin}, M., {et~al.} 2012,
  \href{http://dx.doi.org/10.1088/0004-637X/753/2/163}{\JournalTitle{\apj},
  753, 163}

\bibitem[{{Gonzalez} {et~al.}(2020){Gonzalez}, {Chalela}, {Jauzac}, {Eckert},
  {Schaller}, {Harvey}, {Niemiec}, {Koekemoer}, {Barnes}, {Clowe}, {Connor},
  {Diego}, {Remolina Gonzalez}, \& {Steinhardt}}]{Gonzalez:20}
{Gonzalez}, E.~J., {Chalela}, M., {Jauzac}, M., {et~al.} 2020,
  \href{http://dx.doi.org/10.1093/mnras/staa745}{\JournalTitle{\mnras}, 494,
  349}

\bibitem[{{Gralla} {et~al.}(2011){Gralla}, {Sharon}, {Gladders}, {Marrone},
  {Barrientos}, {Bayliss}, {Bonamente}, {Bulbul}, {Carlstrom}, {Culverhouse},
  {Gilbank}, {Greer}, {Hasler}, {Hawkins}, {Hennessy}, {Joy}, {Koester},
  {Lamb}, {Leitch}, {Miller}, {Mroczkowski}, {Muchovej}, {Oguri}, {Plagge},
  {Pryke}, \& {Woody}}]{Gralla:11}
{Gralla}, M.~B., {Sharon}, K., {Gladders}, M.~D., {et~al.} 2011,
  \href{http://dx.doi.org/10.1088/0004-637X/737/2/74}{\JournalTitle{\apj}, 737,
  74}

\bibitem[{{Habib} {et~al.}(2016){Habib}, {Pope}, {Finkel}, {Frontiere},
  {Heitmann}, {Daniel}, {Fasel}, {Morozov}, {Zagaris}, {Peterka}, {Vishwanath},
  {Luki{\'c}}, {Sehrish}, \& {Liao}}]{Habib:16}
{Habib}, S., {Pope}, A., {Finkel}, H., {et~al.} 2016,
  \href{http://dx.doi.org/10.1016/j.newast.2015.06.003}{\JournalTitle{\na}, 42,
  49}

\bibitem[{{Heitmann} {et~al.}(2019){Heitmann}, {Finkel}, {Pope}, {Morozov},
  {Frontiere}, {Habib}, {Rangel}, {Uram}, {Korytov}, {Child}, {Flender},
  {Insley}, \& {Rizzi}}]{Heitmann:19}
{Heitmann}, K., {Finkel}, H., {Pope}, A., {et~al.} 2019,
  \href{http://dx.doi.org/10.3847/1538-4365/ab4da1}{\JournalTitle{\apjs}, 245,
  16}

\bibitem[{{Hilbert} {et~al.}(2007){Hilbert}, {White}, {Hartlap}, \&
  {Schneider}}]{Hilbert:07}
{Hilbert}, S., {White}, S. D.~M., {Hartlap}, J., \& {Schneider}, P. 2007,
  \href{http://dx.doi.org/10.1111/j.1365-2966.2007.12391.x}{\JournalTitle{\mnras},
  382, 121}

\bibitem[{{Hilbert} {et~al.}(2008){Hilbert}, {White}, {Hartlap}, \&
  {Schneider}}]{Hilbert:08}
---. 2008,
  \href{http://dx.doi.org/10.1111/j.1365-2966.2008.13190.x}{\JournalTitle{\mnras},
  386, 1845}

\bibitem[{{Hilton} {et~al.}(2018){Hilton}, {Hasselfield}, {Sif{\'o}n},
  {Battaglia}, {Aiola}, {Bharadwaj}, {Bond}, {Choi}, {Crichton}, {Datta},
  {Devlin}, {Dunkley}, {D{\"u}nner}, {Gallardo}, {Gralla}, {Hincks}, {Ho},
  {Hubmayr}, {Huffenberger}, {Hughes}, {Koopman}, {Kosowsky}, {Louis},
  {Madhavacheril}, {Marriage}, {Maurin}, {McMahon}, {Miyatake}, {Moodley},
  {N{\ae}ss}, {Nati}, {Newburgh}, {Niemack}, {Oguri}, {Page}, {Partridge},
  {Schmitt}, {Sievers}, {Spergel}, {Staggs}, {Trac}, {van Engelen},
  {Vavagiakis}, \& {Wollack}}]{Hilton:18}
{Hilton}, M., {Hasselfield}, M., {Sif{\'o}n}, C., {et~al.} 2018,
  \href{http://dx.doi.org/10.3847/1538-4365/aaa6cb}{\JournalTitle{The
  Astrophysical Journal Supplement Series}, 235, 20}

\bibitem[{{Hu}(2003)}]{Hu:03b}
{Hu}, W. 2003,
  \href{http://dx.doi.org/10.1103/PhysRevD.67.081304}{\JournalTitle{\prd}, 67,
  081304}

\bibitem[{{Huang} {et~al.}(2020{\natexlab{a}}){Huang}, {Bleem}, {Stalder},
  {Ade}, {Allen}, {Anderson}, {Austermann}, {Avva}, {Beall}, {Bender},
  {Benson}, {Bianchini}, {Bocquet}, {Brodwin}, {Carlstrom}, {Chang}, {Chiang},
  {Citron}, {Moran}, {Crawford}, {Crites}, {Haan}, {Dobbs}, {Everett}, {Floyd},
  {Gallicchio}, {George}, {Gilbert}, {Gladders}, {Guns}, {Gupta}, {Halverson},
  {Harrington}, {Henning}, {Hilton}, {Holder}, {Holzapfel}, {Hrubes},
  {Hubmayr}, {Irwin}, {Khullar}, {Knox}, {Lee}, {Li}, {Lowitz}, {McDonald},
  {McMahon}, {Meyer}, {Mocanu}, {Montgomery}, {Nadolski}, {Natoli}, {Nibarger},
  {Noble}, {Novosad}, {Padin}, {Patil}, {Pryke}, {Reichardt}, {Ruhl},
  {Saliwanchik}, {Saro}, {Sayre}, {Schaffer}, {Sharon}, {Sievers}, {Smecher},
  {Stark}, {Story}, {Tucker}, {Vanderlinde}, {Veach}, {Vieira}, {Wang},
  {Whitehorn}, {Wu}, \& {Yefremenko}}]{HuangN:20}
{Huang}, N., {Bleem}, L.~E., {Stalder}, B., {et~al.} 2020{\natexlab{a}},
  \href{http://dx.doi.org/10.3847/1538-3881/ab6a96}{\JournalTitle{\aj}, 159,
  110}

\bibitem[{{Huang} {et~al.}(2020{\natexlab{b}}){Huang}, {Storfer}, {Gu}, {Ravi},
  {Pilon}, {Sheu}, {Venguswamy}, {Bankda}, {Dey}, {Landriau}, {Lang},
  {Meisner}, {Moustakas}, {Myers}, {Sajith}, {Schlafly}, \&
  {Schlegel}}]{HuangX:20}
{Huang}, X., {Storfer}, C., {Gu}, A., {et~al.} 2020{\natexlab{b}},
  \JournalTitle{arXiv e-prints}, arXiv:2005.04730

\bibitem[{{Huterer} \& {Shafer}(2018)}]{Huterer:18}
{Huterer}, D., \& {Shafer}, D.~L. 2018,
  \href{http://dx.doi.org/10.1088/1361-6633/aa997e}{\JournalTitle{Reports on
  Progress in Physics}, 81, 016901}

\bibitem[{{Jacobs} {et~al.}(2019){Jacobs}, {Collett}, {Glazebrook},
  {Buckley-Geer}, {Diehl}, {Lin}, {McCarthy}, {Qin}, {Odden}, {Caso Escudero},
  {Dial}, {Yung}, {Gaitsch}, {Pellico}, {Lindgren}, {Abbott}, {Annis}, {Avila},
  {Brooks}, {Burke}, {Carnero Rosell}, {Carrasco Kind}, {Carretero}, {da
  Costa}, {De Vicente}, {Fosalba}, {Frieman}, {Garc{\'\i}a-Bellido},
  {Gaztanaga}, {Goldstein}, {Gruen}, {Gruendl}, {Gschwend}, {Hollowood},
  {Honscheid}, {Hoyle}, {James}, {Krause}, {Kuropatkin}, {Lahav}, {Lima},
  {Maia}, {Marshall}, {Miquel}, {Plazas}, {Roodman}, {Sanchez}, {Scarpine},
  {Serrano}, {Sevilla-Noarbe}, {Smith}, {Sobreira}, {Suchyta}, {Swanson},
  {Tarle}, {Vikram}, {Walker}, {Zhang}, \& {DES Collaboration}}]{Jacobs:19}
{Jacobs}, C., {Collett}, T., {Glazebrook}, K., {et~al.} 2019,
  \href{http://dx.doi.org/10.3847/1538-4365/ab26b6}{\JournalTitle{\apjs}, 243,
  17}

\bibitem[{{Jauzac} {et~al.}(2020){Jauzac}, {Klein}, {Kneib}, {Richard},
  {Rexroth}, {Sch{\"a}fer}, \& {Verdier}}]{Jauzac:20}
{Jauzac}, M., {Klein}, B., {Kneib}, J.-P., {et~al.} 2020, \JournalTitle{arXiv
  e-prints}, arXiv:2006.10700

\bibitem[{{Jenkins} {et~al.}(2001){Jenkins}, {Frenk}, {White}, {Colberg},
  {Cole}, {Evrard}, {Couchman}, \& {Yoshida}}]{Jenkins:01}
{Jenkins}, A., {Frenk}, C.~S., {White}, S.~D.~M., {et~al.} 2001,
  \href{http://dx.doi.org/10.1046/j.1365-8711.2001.04029.x}{\JournalTitle{\mnras},
  321, 372}

\bibitem[{{Johnson} \& {Sharon}(2016)}]{Johnson:16}
{Johnson}, T.~L., \& {Sharon}, K. 2016,
  \href{http://dx.doi.org/10.3847/0004-637X/832/1/82}{\JournalTitle{\apj}, 832,
  82}

\bibitem[{{Johnson} {et~al.}(2014){Johnson}, {Sharon}, {Bayliss}, {Gladders},
  {Coe}, \& {Ebeling}}]{Johnson:14}
{Johnson}, T.~L., {Sharon}, K., {Bayliss}, M.~B., {et~al.} 2014,
  \href{http://dx.doi.org/10.1088/0004-637X/797/1/48}{\JournalTitle{\apj}, 797,
  48}

\bibitem[{{Johnson} {et~al.}(2017{\natexlab{a}}){Johnson}, {Sharon},
  {Gladders}, {Rigby}, {Bayliss}, {Wuyts}, {Whitaker}, {Florian}, \&
  {Murray}}]{Johnson:17a}
{Johnson}, T.~L., {Sharon}, K., {Gladders}, M.~D., {et~al.} 2017{\natexlab{a}},
  \href{http://dx.doi.org/10.3847/1538-4357/aa7756}{\JournalTitle{\apj}, 843,
  78}

\bibitem[{{Johnson} {et~al.}(2017{\natexlab{b}}){Johnson}, {Rigby}, {Sharon},
  {Gladders}, {Florian}, {Bayliss}, {Wuyts}, {Whitaker}, {Livermore}, \&
  {Murray}}]{Johnson:17b}
{Johnson}, T.~L., {Rigby}, J.~R., {Sharon}, K., {et~al.} 2017{\natexlab{b}},
  \href{http://dx.doi.org/10.3847/2041-8213/aa7516}{\JournalTitle{\apjl}, 843,
  L21}

\bibitem[{{Jullo} {et~al.}(2007){Jullo}, {Kneib}, {Limousin},
  {El{\'\i}asd{\'o}ttir}, {Marshall}, \& {Verdugo}}]{Jullo:07}
{Jullo}, E., {Kneib}, J.~P., {Limousin}, M., {et~al.} 2007,
  \href{http://dx.doi.org/10.1088/1367-2630/9/12/447}{\JournalTitle{New Journal
  of Physics}, 9, 447}

\bibitem[{{Kawamata} {et~al.}(2016){Kawamata}, {Oguri}, {Ishigaki},
  {Shimasaku}, \& {Ouchi}}]{Kawamata:16}
{Kawamata}, R., {Oguri}, M., {Ishigaki}, M., {Shimasaku}, K., \& {Ouchi}, M.
  2016,
  \href{http://dx.doi.org/10.3847/0004-637X/819/2/114}{\JournalTitle{\apj},
  819, 114}

\bibitem[{{Khedekar} \& {Majumdar}(2013)}]{Khedekar:13}
{Khedekar}, S., \& {Majumdar}, S. 2013,
  \href{http://dx.doi.org/10.1088/1475-7516/2013/02/030}{\JournalTitle{Journal
  of Cosmology and Astro-Particle Physics}, 2013, 030}

\bibitem[{{Khullar} {et~al.}(2020){Khullar}, {Gozman}, {Lin}, {Martinez},
  {Matthews Acu{\~n}a}, {Medina}, {Merz}, {Sanchez}, {Sisco}, {Kavin Stein},
  {Sukay}, {Tavangar}, {Bayliss}, {Bleem}, {Brownsberger}, {Dahle}, {Florian},
  {Gladders}, {Mahler}, {Rigby}, {Sharon}, \& {Stark}}]{Khullar:20}
{Khullar}, G., {Gozman}, K., {Lin}, J.~J., {et~al.} 2020, \JournalTitle{arXiv
  e-prints}, arXiv:2011.06601

\bibitem[{{Komatsu} {et~al.}(2011){Komatsu}, {Smith}, {Dunkley}, {Bennett},
  {Gold}, {Hinshaw}, {Jarosik}, {Larson}, {Nolta}, {Page}, {Spergel},
  {Halpern}, {Hill}, {Kogut}, {Limon}, {Meyer}, {Odegard}, {Tucker}, {Weiland},
  {Wollack}, \& {Wright}}]{Komatsu:11}
{Komatsu}, E., {Smith}, K.~M., {Dunkley}, J., {et~al.} 2011,
  \href{http://dx.doi.org/10.1088/0067-0049/192/2/18}{\JournalTitle{The
  Astrophysical Journal Supplement Series}, 192, 18}

\bibitem[{{Lagattuta} {et~al.}(2019){Lagattuta}, {Richard}, {Bauer},
  {Cl{\'e}ment}, {Mahler}, {Soucail}, {Carton}, {Kneib}, {Laporte}, {Martinez},
  {Patr{\'\i}cio}, {Payne}, {Pell{\'o}}, {Schmidt}, \& {de la
  Vieuville}}]{Lagattuta:19}
{Lagattuta}, D.~J., {Richard}, J., {Bauer}, F.~E., {et~al.} 2019,
  \href{http://dx.doi.org/10.1093/mnras/stz620}{\JournalTitle{\mnras}, 485,
  3738}

\bibitem[{{Laureijs} {et~al.}(2011){Laureijs}, {Amiaux}, {Arduini},
  {Augu{\`e}res}, {Brinchmann}, {Cole}, {Cropper}, {Dabin}, {Duvet}, {Ealet},
  {Garilli}, {Gondoin}, {Guzzo}, {Hoar}, {Hoekstra}, {Holmes}, {Kitching},
  {Maciaszek}, {Mellier}, {Pasian}, {Percival}, {Rhodes}, {Saavedra Criado},
  {Sauvage}, {Scaramella}, {Valenziano}, {Warren}, {Bender}, {Castander},
  {Cimatti}, {Le F{\`e}vre}, {Kurki-Suonio}, {Levi}, {Lilje}, {Meylan},
  {Nichol}, {Pedersen}, {Popa}, {Rebolo Lopez}, {Rix}, {Rottgering},
  {Zeilinger}, {Grupp}, {Hudelot}, {Massey}, {Meneghetti}, {Miller}, {Paltani},
  {Paulin-Henriksson}, {Pires}, {Saxton}, {Schrabback}, {Seidel}, {Walsh},
  {Aghanim}, {Amendola}, {Bartlett}, {Baccigalupi}, {Beaulieu}, {Benabed},
  {Cuby}, {Elbaz}, {Fosalba}, {Gavazzi}, {Helmi}, {Hook}, {Irwin}, {Kneib},
  {Kunz}, {Mannucci}, {Moscardini}, {Tao}, {Teyssier}, {Weller}, {Zamorani},
  {Zapatero Osorio}, {Boulade}, {Foumond}, {Di Giorgio}, {Guttridge}, {James},
  {Kemp}, {Martignac}, {Spencer}, {Walton}, {Bl{\"u}mchen}, {Bonoli},
  {Bortoletto}, {Cerna}, {Corcione}, {Fabron}, {Jahnke}, {Ligori}, {Madrid},
  {Martin}, {Morgante}, {Pamplona}, {Prieto}, {Riva}, {Toledo}, {Trifoglio},
  {Zerbi}, {Abdalla}, {Douspis}, {Grenet}, {Borgani}, {Bouwens}, {Courbin},
  {Delouis}, {Dubath}, {Fontana}, {Frailis}, {Grazian}, {Koppenh{\"o}fer},
  {Mansutti}, {Melchior}, {Mignoli}, {Mohr}, {Neissner}, {Noddle}, {Poncet},
  {Scodeggio}, {Serrano}, {Shane}, {Starck}, {Surace}, {Taylor},
  {Verdoes-Kleijn}, {Vuerli}, {Williams}, {Zacchei}, {Altieri}, {Escudero
  Sanz}, {Kohley}, {Oosterbroek}, {Astier}, {Bacon}, {Bardelli}, {Baugh},
  {Bellagamba}, {Benoist}, {Bianchi}, {Biviano}, {Branchini}, {Carbone},
  {Cardone}, {Clements}, {Colombi}, {Conselice}, {Cresci}, {Deacon}, {Dunlop},
  {Fedeli}, {Fontanot}, {Franzetti}, {Giocoli}, {Garcia-Bellido}, {Gow},
  {Heavens}, {Hewett}, {Heymans}, {Holland}, {Huang}, {Ilbert}, {Joachimi},
  {Jennins}, {Kerins}, {Kiessling}, {Kirk}, {Kotak}, {Krause}, {Lahav}, {van
  Leeuwen}, {Lesgourgues}, {Lombardi}, {Magliocchetti}, {Maguire}, {Majerotto},
  {Maoli}, {Marulli}, {Maurogordato}, {McCracken}, {McLure}, {Melchiorri},
  {Merson}, {Moresco}, {Nonino}, {Norberg}, {Peacock}, {Pello}, {Penny},
  {Pettorino}, {Di Porto}, {Pozzetti}, {Quercellini}, {Radovich}, {Rassat},
  {Roche}, {Ronayette}, {Rossetti}, {Sartoris}, {Schneider}, {Semboloni},
  {Serjeant}, {Simpson}, {Skordis}, {Smadja}, {Smartt}, {Spano}, {Spiro},
  {Sullivan}, {Tilquin}, {Trotta}, {Verde}, {Wang}, {Williger}, {Zhao},
  {Zoubian}, \& {Zucca}}]{Laureijs:11}
{Laureijs}, R., {Amiaux}, J., {Arduini}, S., {et~al.} 2011, \JournalTitle{ArXiv
  e-prints}, arXiv:1110.3193

\bibitem[{{Li} {et~al.}(2016){Li}, {Gladders}, {Rangel}, {Florian}, {Bleem},
  {Heitmann}, {Habib}, \& {Fasel}}]{Li:16}
{Li}, N., {Gladders}, M.~D., {Rangel}, E.~M., {et~al.} 2016,
  \href{http://dx.doi.org/10.3847/0004-637X/828/1/54}{\JournalTitle{\apj}, 828,
  54}

\bibitem[{{Li} {et~al.}(2019){Li}, {Gladders}, {Heitmann}, {Rangel}, {Child},
  {Florian}, {Bleem}, {Habib}, \& {Finkel}}]{Li:19}
{Li}, N., {Gladders}, M.~D., {Heitmann}, K., {et~al.} 2019,
  \href{http://dx.doi.org/10.3847/1538-4357/ab1f74}{\JournalTitle{\apj}, 878,
  122}

\bibitem[{{Lotz} {et~al.}(2017){Lotz}, {Koekemoer}, {Coe}, {Grogin}, {Capak},
  {Mack}, {Anderson}, {Avila}, {Barker}, {Borncamp}, {Brammer}, {Durbin},
  {Gunning}, {Hilbert}, {Jenkner}, {Khandrika}, {Levay}, {Lucas}, {MacKenty},
  {Ogaz}, {Porterfield}, {Reid}, {Robberto}, {Royle}, {Smith},
  {Storrie-Lombardi}, {Sunnquist}, {Surace}, {Taylor}, {Williams}, {Bullock},
  {Dickinson}, {Finkelstein}, {Natarajan}, {Richard}, {Robertson}, {Tumlinson},
  {Zitrin}, {Flanagan}, {Sembach}, {Soifer}, \& {Mountain}}]{Lotz:17}
{Lotz}, J.~M., {Koekemoer}, A., {Coe}, D., {et~al.} 2017,
  \href{http://dx.doi.org/10.3847/1538-4357/837/1/97}{\JournalTitle{\apj}, 837,
  97}

\bibitem[{{LSST Science Collaboration} {et~al.}(2009){LSST Science
  Collaboration}, {Abell}, {Allison}, {Anderson}, {Andrew}, {Angel}, {Armus},
  {Arnett}, {Asztalos}, {Axelrod}, {Bailey}, {Ballantyne}, {Bankert},
  {Barkhouse}, {Barr}, {Barrientos}, {Barth}, {Bartlett}, {Becker}, {Becla},
  {Beers}, {Bernstein}, {Biswas}, {Blanton}, {Bloom}, {Bochanski}, {Boeshaar},
  {Borne}, {Bradac}, {Brandt}, {Bridge}, {Brown}, {Brunner}, {Bullock},
  {Burgasser}, {Burge}, {Burke}, {Cargile}, {Chand rasekharan}, {Chartas},
  {Chesley}, {Chu}, {Cinabro}, {Claire}, {Claver}, {Clowe}, {Connolly}, {Cook},
  {Cooke}, {Cooray}, {Covey}, {Culliton}, {de Jong}, {de Vries}, {Debattista},
  {Delgado}, {Dell'Antonio}, {Dhital}, {Di Stefano}, {Dickinson}, {Dilday},
  {Djorgovski}, {Dobler}, {Donalek}, {Dubois-Felsmann}, {Durech},
  {Eliasdottir}, {Eracleous}, {Eyer}, {Falco}, {Fan}, {Fassnacht}, {Ferguson},
  {Fernandez}, {Fields}, {Finkbeiner}, {Figueroa}, {Fox}, {Francke}, {Frank},
  {Frieman}, {Fromenteau}, {Furqan}, {Galaz}, {Gal-Yam}, {Garnavich},
  {Gawiser}, {Geary}, {Gee}, {Gibson}, {Gilmore}, {Grace}, {Green}, {Gressler},
  {Grillmair}, {Habib}, {Haggerty}, {Hamuy}, {Harris}, {Hawley}, {Heavens},
  {Hebb}, {Henry}, {Hileman}, {Hilton}, {Hoadley}, {Holberg}, {Holman},
  {Howell}, {Infante}, {Ivezic}, {Jacoby}, {Jain}, {R}, {Jedicke}, {Jee},
  {Garrett Jernigan}, {Jha}, {Johnston}, {Jones}, {Juric}, {Kaasalainen},
  {Styliani}, {Kafka}, {Kahn}, {Kaib}, {Kalirai}, {Kantor}, {Kasliwal},
  {Keeton}, {Kessler}, {Knezevic}, {Kowalski}, {Krabbendam}, {Krughoff},
  {Kulkarni}, {Kuhlman}, {Lacy}, {Lepine}, {Liang}, {Lien}, {Lira}, {Long},
  {Lorenz}, {Lotz}, {Lupton}, {Lutz}, {Macri}, {Mahabal}, {Mandelbaum},
  {Marshall}, {May}, {McGehee}, {Meadows}, {Meert}, {Milani}, {Miller},
  {Miller}, {Mills}, {Minniti}, {Monet}, {Mukadam}, {Nakar}, {Neill}, {Newman},
  {Nikolaev}, {Nordby}, {O'Connor}, {Oguri}, {Oliver}, {Olivier}, {Olsen},
  {Olsen}, {Olszewski}, {Oluseyi}, {Padilla}, {Parker}, {Pepper}, {Peterson},
  {Petry}, {Pinto}, {Pizagno}, {Popescu}, {Prsa}, {Radcka}, {Raddick},
  {Rasmussen}, {Rau}, {Rho}, {Rhoads}, {Richards}, {Ridgway}, {Robertson},
  {Roskar}, {Saha}, {Sarajedini}, {Scannapieco}, {Schalk}, {Schindler},
  {Schmidt}, {Schmidt}, {Schneider}, {Schumacher}, {Scranton}, {Sebag},
  {Seppala}, {Shemmer}, {Simon}, {Sivertz}, {Smith}, {Allyn Smith}, {Smith},
  {Spitz}, {Stanford}, {Stassun}, {Strader}, {Strauss}, {Stubbs}, {Sweeney},
  {Szalay}, {Szkody}, {Takada}, {Thorman}, {Trilling}, {Trimble}, {Tyson}, {Van
  Berg}, {Vand en Berk}, {VanderPlas}, {Verde}, {Vrsnak}, {Walkowicz}, {Wand
  elt}, {Wang}, {Wang}, {Warner}, {Wechsler}, {West}, {Wiecha}, {Williams},
  {Willman}, {Wittman}, {Wolff}, {Wood-Vasey}, {Wozniak}, {Young}, {Zentner},
  \& {Zhan}}]{LSST:09}
{LSST Science Collaboration}, {Abell}, P.~A., {Allison}, J., {et~al.} 2009,
  \JournalTitle{arXiv e-prints}, arXiv:0912.0201

\bibitem[{{LSST Science Collaboration} {et~al.}(2017){LSST Science
  Collaboration}, {Marshall}, {Anguita}, {Bianco}, {Bellm}, {Brandt},
  {Clarkson}, {Connolly}, {Gawiser}, {Ivezic}, {Jones}, {Lochner}, {Lund},
  {Mahabal}, {Nidever}, {Olsen}, {Ridgway}, {Rhodes}, {Shemmer}, {Trilling},
  {Vivas}, {Walkowicz}, {Willman}, {Yoachim}, {Anderson}, {Antilogus}, {Angus},
  {Arcavi}, {Awan}, {Biswas}, {Bell}, {Bennett}, {Britt}, {Buzasi},
  {Casetti-Dinescu}, {Chomiuk}, {Claver}, {Cook}, {Davenport}, {Debattista},
  {Digel}, {Doctor}, {Firth}, {Foley}, {Fong}, {Galbany}, {Giampapa}, {Gizis},
  {Graham}, {Grillmair}, {Gris}, {Haiman}, {Hartigan}, {Hawley}, {Hlozek},
  {Jha}, {Johns-Krull}, {Kanbur}, {Kalogera}, {Kashyap}, {Kasliwal}, {Kessler},
  {Kim}, {Kurczynski}, {Lahav}, {Liu}, {Malz}, {Margutti}, {Matheson},
  {McEwen}, {McGehee}, {Meibom}, {Meyers}, {Monet}, {Neilsen}, {Newman},
  {O'Dowd}, {Peiris}, {Penny}, {Peters}, {Poleski}, {Ponder}, {Richards},
  {Rho}, {Rubin}, {Schmidt}, {Schuhmann}, {Shporer}, {Slater}, {Smith},
  {Soares-Santos}, {Stassun}, {Strader}, {Strauss}, {Street}, {Stubbs},
  {Sullivan}, {Szkody}, {Trimble}, {Tyson}, {de Val-Borro}, {Valenti},
  {Wagoner}, {Wood-Vasey}, \& {Zauderer}}]{LSST:17}
{LSST Science Collaboration}, {Marshall}, P., {Anguita}, T., {et~al.} 2017,
  \JournalTitle{ArXiv e-prints}, arXiv:1708.04058

\bibitem[{{Mahler} {et~al.}(2018){Mahler}, {Richard}, {Cl{\'e}ment},
  {Lagattuta}, {Schmidt}, {Patr{\'\i}cio}, {Soucail}, {Bacon}, {Pello},
  {Bouwens}, {Maseda}, {Martinez}, {Carollo}, {Inami}, {Leclercq}, \&
  {Wisotzki}}]{Mahler:18}
{Mahler}, G., {Richard}, J., {Cl{\'e}ment}, B., {et~al.} 2018,
  \href{http://dx.doi.org/10.1093/mnras/stx1971}{\JournalTitle{\mnras}, 473,
  663}

\bibitem[{{Mahler} {et~al.}(2020){Mahler}, {Sharon}, {Gladders}, {Bleem},
  {Bayliss}, {Calzadilla}, {Floyd}, {Khullar}, {McDonald}, {Remolina
  Gonz{\'a}lez}, {Schrabback}, {Stark}, \& {van den Busch}}]{Mahler:20}
{Mahler}, G., {Sharon}, K., {Gladders}, M.~D., {et~al.} 2020,
  \href{http://dx.doi.org/10.3847/1538-4357/ab886b}{\JournalTitle{\apj}, 894,
  150}

\bibitem[{{Mantz} {et~al.}(2014){Mantz}, {Allen}, {Morris}, {Rapetti},
  {Applegate}, {Kelly}, {von der Linden}, \& {Schmidt}}]{Mantz:14}
{Mantz}, A.~B., {Allen}, S.~W., {Morris}, R.~G., {et~al.} 2014,
  \href{http://dx.doi.org/10.1093/mnras/stu368}{\JournalTitle{\mnras}, 440,
  2077}

\bibitem[{{Marriage} {et~al.}(2011){Marriage}, {Acquaviva}, {Ade}, {Aguirre},
  {Amiri}, {Appel}, {Barrientos}, {Battistelli}, {Bond}, {Brown}, {Burger},
  {Chervenak}, {Das}, {Devlin}, {Dicker}, {Bertrand Doriese}, {Dunkley},
  {D{\"u}nner}, {Essinger-Hileman}, {Fisher}, {Fowler}, {Hajian}, {Halpern},
  {Hasselfield}, {Hern{\'a }ndez-Monteagudo}, {Hilton}, {Hilton}, {Hincks},
  {Hlozek}, {Huffenberger}, {Handel Hughes}, {Hughes}, {Infante}, {Irwin},
  {Baptiste Juin}, {Kaul}, {Klein}, {Kosowsky}, {Lau}, {Limon}, {Lin},
  {Lupton}, {Marsden}, {Martocci}, {Mauskopf}, {Menanteau}, {Moodley},
  {Moseley}, {Netterfield}, {Niemack}, {Nolta}, {Page}, {Parker}, {Partridge},
  {Quintana}, {Reese}, {Reid}, {Sehgal}, {Sherwin}, {Sievers}, {Spergel},
  {Staggs}, {Swetz}, {Switzer}, {Thornton}, {Trac}, {Tucker}, {Warne},
  {Wilson}, {Wollack}, \& {Zhao}}]{Marriage:11}
{Marriage}, T.~A., {Acquaviva}, V., {Ade}, P. A.~R., {et~al.} 2011,
  \href{http://dx.doi.org/10.1088/0004-637X/737/2/61}{\JournalTitle{\apj}, 737,
  61}

\bibitem[{{Meneghetti} {et~al.}(2013){Meneghetti}, {Bartelmann}, {Dahle}, \&
  {Limousin}}]{Meneghetti:13}
{Meneghetti}, M., {Bartelmann}, M., {Dahle}, H., \& {Limousin}, M. 2013,
  \href{http://dx.doi.org/10.1007/s11214-013-9981-x}{\JournalTitle{\ssr}, 177,
  31}

\bibitem[{{Meneghetti} {et~al.}(2003){Meneghetti}, {Bartelmann}, \&
  {Moscardini}}]{Meneghetti:03}
{Meneghetti}, M., {Bartelmann}, M., \& {Moscardini}, L. 2003,
  \href{http://dx.doi.org/10.1046/j.1365-2966.2003.07068.x}{\JournalTitle{\mnras},
  346, 67}

\bibitem[{{Meneghetti} {et~al.}(2014){Meneghetti}, {Rasia}, {Vega}, {Merten},
  {Postman}, {Yepes}, {Sembolini}, {Donahue}, {Ettori}, {Umetsu}, {Balestra},
  {Bartelmann}, {Ben{\'\i}tez}, {Biviano}, {Bouwens}, {Bradley}, {Broadhurst},
  {Coe}, {Czakon}, {De Petris}, {Ford}, {Giocoli}, {Gottl{\"o}ber}, {Grillo},
  {Infante}, {Jouvel}, {Kelson}, {Koekemoer}, {Lahav}, {Lemze}, {Medezinski},
  {Melchior}, {Mercurio}, {Molino}, {Moscardini}, {Monna}, {Moustakas},
  {Moustakas}, {Nonino}, {Rhodes}, {Rosati}, {Sayers}, {Seitz}, {Zheng}, \&
  {Zitrin}}]{Meneghetti:14}
{Meneghetti}, M., {Rasia}, E., {Vega}, J., {et~al.} 2014,
  \href{http://dx.doi.org/10.1088/0004-637X/797/1/34}{\JournalTitle{\apj}, 797,
  34}

\bibitem[{{Meneghetti} {et~al.}(2017){Meneghetti}, {Natarajan}, {Coe},
  {Contini}, {De Lucia}, {Giocoli}, {Acebron}, {Borgani}, {Bradac}, {Diego},
  {Hoag}, {Ishigaki}, {Johnson}, {Jullo}, {Kawamata}, {Lam}, {Limousin},
  {Liesenborgs}, {Oguri}, {Sebesta}, {Sharon}, {Williams}, \&
  {Zitrin}}]{Meneghetti:17}
{Meneghetti}, M., {Natarajan}, P., {Coe}, D., {et~al.} 2017,
  \href{http://dx.doi.org/10.1093/mnras/stx2064}{\JournalTitle{\mnras}, 472,
  3177}

\bibitem[{{Merten} {et~al.}(2015){Merten}, {Meneghetti}, {Postman}, {Umetsu},
  {Zitrin}, {Medezinski}, {Nonino}, {Koekemoer}, {Melchior}, {Gruen},
  {Moustakas}, {Bartelmann}, {Host}, {Donahue}, {Coe}, {Molino}, {Jouvel},
  {Monna}, {Seitz}, {Czakon}, {Lemze}, {Sayers}, {Balestra}, {Rosati},
  {Ben{\'\i}tez}, {Biviano}, {Bouwens}, {Bradley}, {Broadhurst}, {Carrasco},
  {Ford}, {Grillo}, {Infante}, {Kelson}, {Lahav}, {Massey}, {Moustakas},
  {Rasia}, {Rhodes}, {Vega}, \& {Zheng}}]{Merten:15}
{Merten}, J., {Meneghetti}, M., {Postman}, M., {et~al.} 2015,
  \href{http://dx.doi.org/10.1088/0004-637X/806/1/4}{\JournalTitle{\apj}, 806,
  4}

\bibitem[{{Mittal} {et~al.}(2018){Mittal}, {de Bernardis}, \&
  {Niemack}}]{Mittal:18}
{Mittal}, A., {de Bernardis}, F., \& {Niemack}, M.~D. 2018,
  \href{http://dx.doi.org/10.1088/1475-7516/2018/02/032}{\JournalTitle{Journal
  of Cosmology and Astro-Particle Physics}, 2018, 032}

\bibitem[{{Nord} {et~al.}(2016){Nord}, {Buckley-Geer}, {Lin}, {Diehl},
  {Helsby}, {Kuropatkin}, {Amara}, {Collett}, {Allam}, {Caminha}, {De Bom},
  {Desai}, {D{\'u}met-Montoya}, {Pereira}, {Finley}, {Flaugher}, {Furlanetto},
  {Gaitsch}, {Gill}, {Merritt}, {More}, {Tucker}, {Saro}, {Rykoff}, {Rozo},
  {Birrer}, {Abdalla}, {Agnello}, {Auger}, {Brunner}, {Carrasco Kind},
  {Castander}, {Cunha}, {da Costa}, {Foley}, {Gerdes}, {Glazebrook},
  {Gschwend}, {Hartley}, {Kessler}, {Lagattuta}, {Lewis}, {Maia}, {Makler},
  {Menanteau}, {Niernberg}, {Scolnic}, {Vieira}, {Gramillano}, {Abbott},
  {Banerji}, {Benoit-L{\'e}vy}, {Brooks}, {Burke}, {Capozzi}, {Carnero Rosell},
  {Carretero}, {D'Andrea}, {Dietrich}, {Doel}, {Evrard}, {Frieman},
  {Gaztanaga}, {Gruen}, {Honscheid}, {James}, {Kuehn}, {Li}, {Lima},
  {Marshall}, {Martini}, {Melchior}, {Miquel}, {Neilsen}, {Nichol}, {Ogando},
  {Plazas}, {Romer}, {Sako}, {Sanchez}, {Scarpine}, {Schubnell},
  {Sevilla-Noarbe}, {Smith}, {Soares-Santos}, {Sobreira}, {Suchyta}, {Swanson},
  {Tarle}, {Thaler}, {Walker}, {Wester}, \& {Zhang}}]{Nord:16}
{Nord}, B., {Buckley-Geer}, E., {Lin}, H., {et~al.} 2016,
  \href{http://dx.doi.org/10.3847/0004-637X/827/1/51}{\JournalTitle{\apj}, 827,
  51}

\bibitem[{{Nord} {et~al.}(2020){Nord}, {Buckley-Geer}, {Lin}, {Kuropatkin},
  {Collett}, {Tucker}, {Diehl}, {Agnello}, {Amara}, {Abbott}, {Allam}, {Annis},
  {Avila}, {Bechtol}, {Brooks}, {Burke}, {Carnero Rosell}, {Carrasco Kind},
  {Carretero}, {Cunha}, {da Costa}, {Davis}, {De Vicente}, {Doel}, {Eifler},
  {Evrard}, {Fernandez}, {Flaugher}, {Fosalba}, {Frieman},
  {Garc{\'\i}a-Bellido}, {Gaztanaga}, {Gruen}, {Gruendl}, {Gutierrez},
  {Hartley}, {Hollowood}, {Honscheid}, {Hoyle}, {James}, {Kuehn}, {Lahav},
  {Lima}, {Maia}, {March}, {Marshall}, {Melchior}, {Menanteau}, {Miquel},
  {Plazas}, {Romer}, {Roodman}, {Rykoff}, {Sanchez}, {Scarpine}, {Schindler},
  {Schubnell}, {Sevilla-Noarbe}, {Smith}, {Soares-Santos}, {Sobreira},
  {Suchyta}, {Swanson}, {Tarle}, {Thomas}, {Zhang}, \& {DES
  Collaboration}}]{Nord:20}
---. 2020,
  \href{http://dx.doi.org/10.1093/mnras/staa200}{\JournalTitle{\mnras}, 494,
  1308}

\bibitem[{{Oguri}(2006)}]{Oguri:06}
{Oguri}, M. 2006,
  \href{http://dx.doi.org/10.1111/j.1365-2966.2006.10043.x}{\JournalTitle{\mnras},
  367, 1241}

\bibitem[{{Oguri} {et~al.}(2012){Oguri}, {Bayliss}, {Dahle}, {Sharon},
  {Gladders}, {Natarajan}, {Hennawi}, \& {Koester}}]{Oguri:12}
{Oguri}, M., {Bayliss}, M.~B., {Dahle}, H., {et~al.} 2012,
  \href{http://dx.doi.org/10.1111/j.1365-2966.2011.20248.x}{\JournalTitle{\mnras},
  420, 3213}

\bibitem[{{Oguri} \& {Blandford}(2009)}]{Oguri:09}
{Oguri}, M., \& {Blandford}, R.~D. 2009,
  \href{http://dx.doi.org/10.1111/j.1365-2966.2008.14154.x}{\JournalTitle{\mnras},
  392, 930}

\bibitem[{{Paterno-Mahler} {et~al.}(2018){Paterno-Mahler}, {Sharon}, {Coe},
  {Mahler}, {Cerny}, {Johnson}, {Schrabback}, {Andrade-Santos}, {Avila},
  {Brada{\v{c}}}, {Bradley}, {Carrasco}, {Czakon}, {Dawson}, {Frye}, {Hoag},
  {Huang}, {Jones}, {Lam}, {Livermore}, {Lovisari}, {Mainali}, {Oesch}, {Ogaz},
  {Past}, {Peterson}, {Ryan}, {Salmon}, {Sendra-Server}, {Stark}, {Umetsu},
  {Vulcani}, \& {Zitrin}}]{Paterno-Mahler:18}
{Paterno-Mahler}, R., {Sharon}, K., {Coe}, D., {et~al.} 2018,
  \href{http://dx.doi.org/10.3847/1538-4357/aad239}{\JournalTitle{\apj}, 863,
  154}

\bibitem[{{Pearson} {et~al.}(2019){Pearson}, {Li}, \& {Dye}}]{Pearson:19}
{Pearson}, J., {Li}, N., \& {Dye}, S. 2019,
  \href{http://dx.doi.org/10.1093/mnras/stz1750}{\JournalTitle{\mnras}, 488,
  991}

\bibitem[{{Petrillo} {et~al.}(2017){Petrillo}, {Tortora}, {Chatterjee},
  {Vernardos}, {Koopmans}, {Verdoes Kleijn}, {Napolitano}, {Covone},
  {Schneider}, {Grado}, \& {McFarland}}]{Petrillo:17}
{Petrillo}, C.~E., {Tortora}, C., {Chatterjee}, S., {et~al.} 2017,
  \href{http://dx.doi.org/10.1093/mnras/stx2052}{\JournalTitle{\mnras}, 472,
  1129}

\bibitem[{{Pillepich} {et~al.}(2018){Pillepich}, {Reiprich}, {Porciani},
  {Borm}, \& {Merloni}}]{Pillepich:18}
{Pillepich}, A., {Reiprich}, T.~H., {Porciani}, C., {Borm}, K., \& {Merloni},
  A. 2018,
  \href{http://dx.doi.org/10.1093/mnras/sty2240}{\JournalTitle{\mnras}, 481,
  613}

\bibitem[{{Pratt} {et~al.}(2019){Pratt}, {Arnaud}, {Biviano}, {Eckert},
  {Ettori}, {Nagai}, {Okabe}, \& {Reiprich}}]{Pratt:19}
{Pratt}, G.~W., {Arnaud}, M., {Biviano}, A., {et~al.} 2019,
  \href{http://dx.doi.org/10.1007/s11214-019-0591-0}{\JournalTitle{\ssr}, 215,
  25}

\bibitem[{{Priewe} {et~al.}(2017){Priewe}, {Williams}, {Liesenborgs}, {Coe}, \&
  {Rodney}}]{Priewe:17}
{Priewe}, J., {Williams}, L. L.~R., {Liesenborgs}, J., {Coe}, D., \& {Rodney},
  S.~A. 2017,
  \href{http://dx.doi.org/10.1093/mnras/stw2785}{\JournalTitle{\mnras}, 465,
  1030}

\bibitem[{{Raney} {et~al.}(2020{\natexlab{a}}){Raney}, {Keeton}, \&
  {Brennan}}]{Raney:20a}
{Raney}, C.~A., {Keeton}, C.~R., \& {Brennan}, S. 2020{\natexlab{a}},
  \href{http://dx.doi.org/10.1093/mnras/stz3116}{\JournalTitle{\mnras}, 492,
  503}

\bibitem[{{Raney} {et~al.}(2020{\natexlab{b}}){Raney}, {Keeton}, {Brennan}, \&
  {Fan}}]{Raney:20b}
{Raney}, C.~A., {Keeton}, C.~R., {Brennan}, S., \& {Fan}, H.
  2020{\natexlab{b}},
  \href{http://dx.doi.org/10.1093/mnras/staa921}{\JournalTitle{\mnras}, 494,
  4771}

\bibitem[{Rangel {et~al.}(2016)Rangel, Li, Habib, Peterka, Agrawal, Liao, \&
  Choudhary}]{Rangel:16}
Rangel, E., Li, N., Habib, S., {et~al.} 2016,
  \href{http://dx.doi.org/10.1109/CLUSTER.2016.40}{in 2016 IEEE International
  Conference on Cluster Computing (CLUSTER)}, 30

\bibitem[{{Remolina Gonz{\'a}lez} {et~al.}(in preparation){Remolina
  Gonz{\'a}lez}, {Sharon}, \& et~al.}]{Remolina:21}
{Remolina Gonz{\'a}lez}, J.~D., {Sharon}, K., \& et~al. in preparation,
  \JournalTitle{in preparation}

\bibitem[{{Remolina Gonz{\'a}lez} {et~al.}(2018){Remolina Gonz{\'a}lez},
  {Sharon}, \& {Mahler}}]{Remolina:18}
{Remolina Gonz{\'a}lez}, J.~D., {Sharon}, K., \& {Mahler}, G. 2018,
  \href{http://dx.doi.org/10.3847/1538-4357/aacf8e}{\JournalTitle{\apj}, 863,
  60}

\bibitem[{{Remolina Gonz{\'a}lez} {et~al.}(2020){Remolina Gonz{\'a}lez},
  {Sharon}, {Reed}, {Li}, {Mahler}, {Bleem}, {Gladders}, {Niemiec}, {Acebron},
  \& {Child}}]{Remolina:20}
{Remolina Gonz{\'a}lez}, J.~D., {Sharon}, K., {Reed}, B., {et~al.} 2020,
  \href{http://dx.doi.org/10.3847/1538-4357/abb2a1}{\JournalTitle{\apj}, 902,
  44}

\bibitem[{{Richard} {et~al.}(2011){Richard}, {Kneib}, {Ebeling}, {Stark},
  {Egami}, \& {Fiedler}}]{Richard:11}
{Richard}, J., {Kneib}, J.-P., {Ebeling}, H., {et~al.} 2011,
  \href{http://dx.doi.org/10.1111/j.1745-3933.2011.01050.x}{\JournalTitle{\mnras},
  414, L31}

\bibitem[{{Rigby} {et~al.}(2018){Rigby}, {Bayliss}, {Sharon}, {Gladders},
  {Chisholm}, {Dahle}, {Johnson}, {Paterno-Mahler}, {Wuyts}, \&
  {Kelson}}]{Rigby:18}
{Rigby}, J.~R., {Bayliss}, M.~B., {Sharon}, K., {et~al.} 2018,
  \href{http://dx.doi.org/10.3847/1538-3881/aaa2ff}{\JournalTitle{\aj}, 155,
  104}

\bibitem[{{Sebesta} {et~al.}(2019){Sebesta}, {Williams}, {Liesenborgs},
  {Medezinski}, \& {Okabe}}]{Sebesta:19}
{Sebesta}, K., {Williams}, L. L.~R., {Liesenborgs}, J., {Medezinski}, E., \&
  {Okabe}, N. 2019,
  \href{http://dx.doi.org/10.1093/mnras/stz1950}{\JournalTitle{\mnras}, 488,
  3251}

\bibitem[{{Sharon} {et~al.}(2020){Sharon}, {Bayliss}, {Dahle}, {Dunham},
  {Florian}, {Gladders}, {Johnson}, {Mahler}, {Paterno-Mahler}, {Rigby},
  {Whitaker}, {Akhshik}, {Koester}, {Murray}, {Remolina Gonz{\'a}lez}, \&
  {Wuyts}}]{Sharon:20}
{Sharon}, K., {Bayliss}, M.~B., {Dahle}, H., {et~al.} 2020,
  \href{http://dx.doi.org/10.3847/1538-4365/ab5f13}{\JournalTitle{\apjs}, 247,
  12}

\bibitem[{{Shin} {et~al.}(2019){Shin}, {Adhikari}, {Baxter}, {Chang}, {Jain},
  {Battaglia}, {Bleem}, {Bocquet}, {DeRose}, {Gruen}, {Hilton}, {Kravtsov},
  {McClintock}, {Rozo}, {Rykoff}, {Varga}, {Wechsler}, {Wu}, {Zhang}, {Aiola},
  {Allam}, {Bechtol}, {Benson}, {Bertin}, {Bond}, {Brodwin}, {Brooks},
  {Buckley-Geer}, {Burke}, {Carlstrom}, {Carnero Rosell}, {Carrasco Kind},
  {Carretero}, {Castander}, {Choi}, {Cunha}, {Crawford}, {da Costa}, {De
  Vicente}, {Desai}, {Devlin}, {Dietrich}, {Doel}, {Dunkley}, {Eifler},
  {Evrard}, {Flaugher}, {Fosalba}, {Gallardo}, {Garc{\'\i}a-Bellido},
  {Gaztanaga}, {Gerdes}, {Gralla}, {Gruendl}, {Gschwend}, {Gupta}, {Gutierrez},
  {Hartley}, {Hill}, {Ho}, {Hollowood}, {Honscheid}, {Hoyle}, {Huffenberger},
  {Hughes}, {James}, {Jeltema}, {Kim}, {Krause}, {Kuehn}, {Lahav}, {Lima},
  {Madhavacheril}, {Maia}, {Marshall}, {Maurin}, {McMahon}, {Menanteau},
  {Miller}, {Miquel}, {Mohr}, {Naess}, {Nati}, {Newburgh}, {Niemack}, {Ogando},
  {Page}, {Partridge}, {Patil}, {Plazas}, {Rapetti}, {Reichardt}, {Romer},
  {Sanchez}, {Scarpine}, {Schindler}, {Serrano}, {Smith}, {Smith},
  {Soares-Santos}, {Sobreira}, {Staggs}, {Stark}, {Stein}, {Suchyta},
  {Swanson}, {Tarle}, {Thomas}, {van Engelen}, {Wollack}, \& {Xu}}]{Shin:19}
{Shin}, T., {Adhikari}, S., {Baxter}, E.~J., {et~al.} 2019,
  \href{http://dx.doi.org/10.1093/mnras/stz1434}{\JournalTitle{\mnras}, 487,
  2900}

\bibitem[{{Strait} {et~al.}(2018){Strait}, {Brada{\v{c}}}, {Hoag}, {Huang},
  {Treu}, {Wang}, {Amorin}, {Castellano}, {Fontana}, {Lemaux}, {Merlin},
  {Schmidt}, {Schrabback}, {Tomczack}, {Trenti}, \& {Vulcani}}]{Strait:18}
{Strait}, V., {Brada{\v{c}}}, M., {Hoag}, A., {et~al.} 2018,
  \href{http://dx.doi.org/10.3847/1538-4357/aae834}{\JournalTitle{\apj}, 868,
  129}

\bibitem[{{Umetsu} \& {Diemer}(2017)}]{Umetsu:17}
{Umetsu}, K., \& {Diemer}, B. 2017,
  \href{http://dx.doi.org/10.3847/1538-4357/aa5c90}{\JournalTitle{\apj}, 836,
  231}

\bibitem[{{Verdugo} {et~al.}(2011){Verdugo}, {Motta}, {Mu{\~n}oz}, {Limousin},
  {Cabanac}, \& {Richard}}]{Verdugo:11}
{Verdugo}, T., {Motta}, V., {Mu{\~n}oz}, R.~P., {et~al.} 2011,
  \href{http://dx.doi.org/10.1051/0004-6361/201014965}{\JournalTitle{\aap},
  527, A124}

\bibitem[{{Wambsganss} {et~al.}(2004){Wambsganss}, {Bode}, \&
  {Ostriker}}]{Wambsganss:04}
{Wambsganss}, J., {Bode}, P., \& {Ostriker}, J.~P. 2004,
  \href{http://dx.doi.org/10.1086/421459}{\JournalTitle{\apj}, 606, L93}

\bibitem[{{Wambsganss} {et~al.}(2008){Wambsganss}, {Ostriker}, \&
  {Bode}}]{Wambsganss:08}
{Wambsganss}, J., {Ostriker}, J.~P., \& {Bode}, P. 2008,
  \href{http://dx.doi.org/10.1086/527529}{\JournalTitle{\apj}, 676, 753}

\bibitem[{{Zitrin} {et~al.}(2014){Zitrin}, {Zheng}, {Broadhurst}, {Moustakas},
  {Lam}, {Shu}, {Huang}, {Diego}, {Ford}, {Lim}, {Bauer}, {Infante}, {Kelson},
  \& {Molino}}]{Zitrin:14}
{Zitrin}, A., {Zheng}, W., {Broadhurst}, T., {et~al.} 2014,
  \href{http://dx.doi.org/10.1088/2041-8205/793/1/L12}{\JournalTitle{\apjl},
  793, L12}

\end{thebibliography}

\end{document}